\documentclass[preprint2]{aastex}

\newcommand{\new}[1]{{#1}}

\shorttitle{Model for Co-evolution of QSOs and Spheroids}

\begin{document}


\title{A Physical Model for the Co-evolution of QSOs and of their Spheroidal Hosts}


\author{Gian Luigi Granato\altaffilmark{1,2}}
\email{granato@pd.astro.it}
\author{Gianfranco De Zotti\altaffilmark{1,2}}
\email{dezotti@pd.astro.it}
\author{Laura Silva\altaffilmark{3}}
\email{silva@ts.astro.it}
\author{Alessandro Bressan\altaffilmark{1,2}}
\email{bressan@pd.astro.it}
\and
\author{Luigi Danese\altaffilmark{2}}
\email{danese@sissa.it} \altaffiltext{1}{INAF - Osservatorio
Astronomico di Padova,
                 Vicolo Osservatorio, I-35100 Padova, Italy}
\altaffiltext{2}{International School for Advanced Studies,
SISSA/ISAS, Via Beirut 2-4, I-34014 Trieste, Italy}
\altaffiltext{3}{INAF - Osservatorio Astronomico di Trieste, Via
G.B. Tiepolo 11, I-34131 Trieste, Italy}


\begin{abstract}

We present a physically motivated model for the early co-evolution
of massive spheroidal galaxies and active nuclei at their centers.
Within dark matter halos, forming at the rate predicted by the
canonical hierarchical clustering scenario, the gas evolution is
controlled by gravity, radiative cooling, and heating by feedback
from supernovae and from the growing active nucleus. Supernova
heating is increasingly effective with decreasing binding energy
in slowing down the star formation and in driving gas outflows.
The more massive proto-galaxies virializing at earlier times are
thus the sites of the faster star-formation. The correspondingly
higher radiation drag fastens the angular momentum loss by the
gas, resulting in a larger accretion rate onto the central
black-hole. In turn, the kinetic energy carried by outflows driven
by active nuclei can unbind the residual gas, thus halting both
the star formation and the black-hole growth, in a time again
shorter for larger halos. For the most massive galaxies the gas
unbinding time is short enough for the bulk of the star-formation
to be completed before type Ia supernovae can substantially
increase the $Fe$ abundance of the interstellar medium, thus
accounting for the $\alpha$-enhancement seen in the largest
galaxies. The feedback from supernovae and from the active nucleus
also determines the relationship between the black-hole mass and
the mass, or the velocity dispersion, of the host galaxy, as well
as the black-hole mass function. In both cases the model
predictions are in excellent agreement with the observational
data. Coupling the model with GRASIL (Silva et al. 1998), the code
computing in a self-consistent way the chemical and
spectrophotometric evolution of galaxies over a very wide
wavelength interval, we have obtained predictions in excellent
agreement with observations for a number of observables which
proved to be extremely challenging for all the current
semi-analytic models, including the sub-mm counts and the
corresponding redshift distributions, and the epoch-dependent
K-band luminosity function of spheroidal galaxies.
\end{abstract}






\section{Introduction}

Although the traditional approach to galaxy formation and
evolution regards nuclear activity as an incidental diversion, it
is becoming clear, beyond any reasonable doubt, that the formation
of super-massive black holes powering nuclear activity is
intimately linked to the formation of its host galaxy and plays a
key role in shaping its evolution. Evidences supporting this view
include: the discovery that Massive Dark Objects (MDOs), with
masses in the range $\sim10^6$--$3\times10^9$~M$_\odot$ and a mass
function matching that of baryons accreted onto black holes during
the quasar activity (Salucci et al. 1999), are present in
essentially all local galaxies with a substantial spheroidal
component (Kormendy \& Richstone 1995; Magorrian et al. 1998; van
der Marel 1999; Kormendy \& Gebhardt 2001); the tight correlation
between the MDO mass (M$_{\rm BH}$) and the velocity dispersion of
stars in the host galaxy (Magorrian et al 1998; Ferrarese \&
Merritt 2000; Gebhardt et al. 2000; Tremaine et al. 2002), the
mass of the spheroidal component (McLure \& Dunlop 2002; Dunlop et
al. 2003; Marconi \& Hunt 2003), and the mass of the dark halo
(Ferrarese 2002); the correspondence between the luminosity
function of active star-forming galaxies at $z\simeq 3$, the
B-band luminosity function of quasars and the mass function of
dark halos at the same redshift (Haehnelt et al. 1998, Monaco et
al. 2000); the similarity between the evolutionary histories of
the luminosity densities of galaxies and quasars (e.g.\ Cavaliere
\& Vittorini 1998). Recently Shields et al. (2003) have found that
the correlation between M$_{\rm BH}$ and the stellar velocity
dispersion is already present at redshift up to $\sim 3$.

As discussed by Granato et al. (2001), the mutual feedback between
galaxies and quasars during their early evolutionary stages may be
the key to overcome some of the crises of the currently standard
scenario for galaxy evolution. For example, predictions of
semi-analytic models (Devriendt \& Guiderdoni 2000; Cole et al.
2000; Somerville et al. 2001; Menci et al. 2002) are persistently
unable to account for the surface density of massive galaxies at
substantial redshift detected by (sub)-mm surveys with SCUBA and
MAMBO (Blain et al. 2002; Scott et al. 2002) and by deep K-band
surveys (Cimatti et al. 2002b; Kashikawa et al. 2003), unless
ad-hoc adjustments are introduced. The difficulty stems from the
fact that the standard CDM scenario tends to imply that most of
the star formation occurs in relatively small galaxies that later
merge to make bigger and bigger objects. On the contrary, the data
indicate that galaxies detected by (sub)-mm surveys are mostly
very massive, with very high star-formation rates ($\sim
10^3\,$M$_\odot\,\hbox{yr}^{-1}$), at $z > 2$ (Dunlop 2001; Ivison
et al. 2002; Aretxaga et al. 2003; Chapman et al. 2003). All these
data are more consistent with the traditional ``monolithic"
scenario, according to which elliptical galaxies formed most of
their stars in a single burst, at relatively high redshifts, and
underwent essentially passive evolution thereafter. On the other
hand, the ``monolithic" scheme is inadequate to the extent that it
cannot be fitted in a consistent scenario for structure formation
from primordial density perturbations.

Clues on the timing of evolution of both galaxies and quasars are
provided by chemical abundances (e.g.\ Fria\c{c}a \& Terlevich
1998). Spectroscopic observations demonstrate, even for the
highest redshift quasars, a fast metal enrichment of the
circum-nuclear gas (e.g. Hamann \& Ferland 1999; Fan et al. 2000,
2001; Freudling et al. 2003). Statistical studies of local E/S0
galaxies, hosting super-massive black-holes, show that the most
massive galaxies are also the most metal rich, the reddest, and,
perhaps, the oldest (Forbes \& Ponman 1999; Trager et al.
2000a,b). Such galaxies show an excess $\alpha$-elements/$Fe$
ratio, compared to solar ($\alpha$-enhancement; Trager et al.
2000a,b; Thomas et al. 2002), suggestive of very intense but short
star-formation activity. Also observations of hosts of high
redshift quasars show that they are as massive as expected from
the local $L_{\rm QSO}$-$L_{\rm host}$ relation, and that most of
their stars are relatively old, although star forming regions are
present (Kukula et al. 2001; Hutchings et al. 2002; Ridgway et al.
2002; Stockton \& Ridgway 2001).

The work carried out by our group in the last several years, aimed
at constructing physically grounded models for the joint formation
and evolution of quasars and spheroidal galaxies in the framework
of the standard hierarchical clustering scenario (Granato et al.
2001), has shown that, {\new to account for the surface density of
massive galaxies at substantial redshifts detected by sub-mm SCUBA
surveys (which turns out to be far in excess of predisctions of
standard semi-analytic models), allowing for the observed
relationships between quasars and galaxies,} it is necessary to
assume that the formation of stars and of the central black hole
took place on shorter timescales within more massive dark matter
halos. In other words, the canonical hierarchical CDM scheme -
small clumps collapse first - is reversed for baryon collapse and
the formation of luminous objects ({\it Anti-hierarchical Baryon
Collapse} scenario). This behavior, attributed to the feedback
from supernova explosions and, for the most massive galaxies, from
nuclear activity, may account simultaneously for evolutionary
properties of quasars and of massive spheroidal galaxies (Monaco
et al. 2000; Granato et al. 2001), for clustering properties of
SCUBA galaxies (Magliocchetti et al. 2002; Perrotta et al. 2003),
and for the metal abundances in spheroidal galaxies and in bulges
of later Hubble types hosting super-massive black holes (Romano et
al. 2002).

Recently, Cattaneo \& Bernardi (2003), using the early type galaxy
sample in the Sloan Digital Sky Survey (Bernardi et al.\ 2003a),
investigated the hypothesis that quasars formed together with the
stellar population of ellipticals, finding a consistency with the
observed luminosity function of optical quasars.

However, there is not, as yet, a clear understanding of the
physical mechanisms governing the interactions among the active
nucleus and the host galaxy. Silk \& Rees (1998) and Fabian (1999)
proposed that the relationship between the black-hole mass and the
velocity dispersion (or mass) of the host stellar spheroid may be
the effect of the quasar feedback. However, in its present form,
the proposed mechanism does not imply shorter star-formation times
for more massive galaxies and therefore cannot easily explain the
$\alpha$-enhancement of more massive objects. Wang \& Biermann
(1998) suggested that the formation of both elliptical galaxies
and super-massive black-holes at their centers is related to the
merging of two proto-disks. Kauffmann \& Haenhelt (2000) and
Haenhelt \& Kauffmann (2000) have analyzed the evolution of active
nuclei and of host galaxies in the framework of the hierarchical
clustering scenario, using a semi-analytic approach. In their
scheme, the merging process determines both the evolution of
galaxies and the growth of the black-holes at their centers. This
model predicts a rapid evolution of galaxies with redshift.
Volonteri et al. (2002, 2003) presented a model in which most of
the mass in BHs is assembled in accretion episodes triggered by
merging. They found that the galaxy merging would leave about 10\%
of massive black-holes distributed in galactic halos and a similar
fraction of binary super-massive black-holes in galactic centers.
\new{Di Matteo et al.\ (2003) considered the gas content in
galaxies, as predicted by cosmological hydrodynamical simulations
including sub-grid prescriptions for gas cooling and star
formation (but not for BH growth). They pointed out that the
observed $M_{\rm BH}$-$\sigma$ correlation is well reproduced,
provided that a linear relationship between the gas and BH masses
(at $z>1$) is assumed.}

In this paper we carry out an investigation of the physical
processes driving the growth of the central black hole and the
effect of the energy released by the active nucleus on the
surrounding proto-galactic gas. The corresponding prescriptions
for matter outflow, gas cooling and collapse, star formation, etc.
are implemented in a model for the formation and evolution of
galaxies, interfaced with our code computing their spectral energy
distribution as a function of their age, taking into account the
evolution of stellar populations, of metal abundances, of the dust
content and its distribution (GRASIL: Silva et al. 1998, Silva
1999, Granato et al. 2000). Our aim is to put on a more solid
physical basis the approach by Granato et al. (2001), which partly
relies on empirical recipes.

The model has been tested against the observed correlation of BH
masses with the mass and velocity dispersion of the host galaxies,
the observed K-band and sub-millimeter counts and the associated
redshift distributions, along with the chemical and photometric
properties of the local E/S0 galaxies. In subsequent papers the
model will be used to predict the time-dependent luminosity
function of quasars up to $z>6$ and of passively evolving
ellipticals, and their clustering properties, with reference to
the observational capabilities of space missions such as SIRTF, to
be launched in a few months, JWST, of ground millimeter telescopes
such as ALMA, and of the forthcoming deep near-IR surveys from the
ground.

In Sect.~2 we describe the basic physical processes governing the
evolution of the star-formation rate in spheroidal galaxies, the
growth of central black-holes and their feedback. In Sect.~3 we
present our main results, which are discussed in Sect. 4. In
Sect.~5 we summarize our main conclusions.

We adopt a cold dark matter cosmology with cosmological constant,
consistent with the Wilkinson Microwave Anisotropy Probe (WMAP)
data (Bennett et al. 2003), as well as with information from large
scale structure (Spergel et al. 2003): $\Omega_m=0.29$,
$\Omega_b=0.047$, and $\Omega_\Lambda=0.71$, $H_0=72~{\rm
km~s^{-1}}$, $\sigma_8=0.8$, and an index $n=1.0$ for the power
spectrum of primordial density fluctuations.

\section{Basic ingredients}

\subsection{Dark matter halos}

Following Navarro et al. (1997) and Bullock et al. (2001b) we
identify as a virialized halo at redshift $z$ a volume of the
Universe of radius $r_{\rm vir}$ enclosing an overdensity
$\Delta_{\rm vir}(z)$ which, for a flat cosmology, can be
approximated by
\begin{equation}
\Delta_{\rm vir} \simeq \frac{(18 \pi^2 +82x-39x^2)}{\Omega(z)},
\label{overdensity}
\end{equation}
where $x= \Omega(z)-1$ and $\Omega(z)$ is the ratio of the mean
matter density to the critical density at redshift $z$ (Bryan \&
Norman 1998). The halo mass is then given by
\begin{equation}
M_{\rm vir}=\frac{4 \pi}{3} \Delta_{\rm vir}(z)\rho_{u}(z) r_{\rm
vir}^3, \label{Mvir}
\end{equation}
$\rho_{u}(z)$ being the mean universal matter density. Useful
quantities are the rotational velocity of the DM halo at its
virial radius, $V^2_{\rm vir} = G M_{\rm vir}/r_{\rm vir}$, and
the equilibrium gas temperature in the DM potential well
\begin{equation}
k \, T=\frac{1}{2} \mu m_{p} V_{\rm vir}^2, \label{Tvir}
\end{equation}
where $m_{p}$ is the proton mass and $\mu m_{p}$ is the mean
molecular weight of the gas.

The virialized halos exhibit a universal density profile (Navarro
et al. 1997;  Bullock et al. 2001b) well described by
\begin{equation}
\rho(r) = \frac{\rho_s}{c\,x(1 + c\,x)^2},
\end{equation}
where $x = r/r_{\rm vir}$ and $c$ is the concentration parameter.
Analyzing the profiles of a large sample of virialized halos
obtained through high-resolution N-body simulations, Bullock et
al. (2001b) found that, at any redshift, $c$ is a weak function of
the mass $c\propto M_{\rm vir}^{-0.13}$, while, at fixed mass,
$c\propto (1+z)^{-1}$.

Numerical simulations and theoretical arguments show that the dark
matter assembly in halos proceeds through a first phase of fast
accretion, followed by a slow phase (Wechsler et al. 2002; van den
Bosch 2002). Zhao et al. (2003) showed that the potential well of
the halos is built up during the fast accretion. \new{The
subsequent slow accretion does not change the ``identity'' of the
halo, though significantly increasing its mass.} Guided by these
results, in the following we will assume that, for the mass and
redshift ranges we are interested in, the virialization epoch of
DM halos coincides with the end of the fast accretion phase and
with the beginning of the vigorous star formation of the
proto-spheroidal galaxies, and that these galaxies keep their
identity till the present time. Our approach implies, in the mass
and redshift range considered, a one to one correspondence between
halos and galaxies, consistent with the available data on
clustering of Lyman-break (Bullock et al. 2002) and SCUBA galaxies
(Magliocchetti et al. 2001; Perrotta et al. 2003).

The formation rate of massive halos ($M_{\rm vir} \gtrsim 2.5
\times 10^{11}\, M_\odot$) at $z\gtrsim 1.5$ is approximated by
the positive part of the time derivative of the halo mass function
$n(M_{\rm vir},z)$ (Haehnelt \& Rees 1993; Sasaki 1994; Peacock
1999). The negative part is small for this mass and redshift
range, consistent with our assumption that massive halos survive
till the present time. Following Press \& Schechter (1974) we
write:
\begin{equation}
n(M_{\rm vir},z)=\frac {\rho} {M_{\rm vir}^2} \nu f(\nu ) \frac
{d\ln \nu } {d \ln M_{\rm vir}}
\end{equation}
where $\rho$ is the average comoving density of the universe and
$\nu= [\delta_c(z)/\sigma_\delta(M_{\rm vir})]^2$, $\sigma_\delta$
being the rms initial density fluctuations smoothed on a scale
containing a mass $M_{\rm vir}$, and $\delta_c$ the critical
over-density for spherical collapse. The latter quantity is given
by $\delta_c(z)=\delta_0/ b(z)$, $\delta_0$ being the present day
critical over-density (1.686 for an Einstein--de Sitter cosmology,
with only negligible dependencies on $\Omega_m$ and
$\Omega_{\Lambda}$) and $b(z)$ the linear growth factor, for which
we adopt the approximation proposed by Li \& Ostriker (2002). For
the function $\nu f(\nu)$ we adopt the expression given by Sheth
\& Tormen (2002), which also takes into account the effect of
ellipsoidal collapse:
\begin{equation}
\nu f(\nu) =A[1+(a\nu)^{-p}](\frac {a\nu}{2})^{1/2}\frac{e^{-a\nu
/2}}{\pi ^{1/2}},
\end{equation}
where $A=0.322$, $p=0.3$ and $a=0.707$.

\subsection{Star-formation rate}

Recent studies on the angular momentum of dark matter (DM) halos
suggest that a significant fraction of the mass has a low specific
angular momentum (Bullock et al. 2001a). Also, the results of high
resolution simulations including gas suggest that a significant
fraction of the angular momentum of the baryons is redistributed
to the dark matter by dynamical friction (see e.g. Navarro \&
Steinmetz 2000), although the heating of the gas in sub-galactic
halos may increase the tidal stripping and the angular momentum of
the gas (Maller \& Dekel 2002). As far as the effect of angular
momentum can be neglected, as is probably the case for the
formation of spheroidal galaxies, the collapse time of baryons
within the host dark matter halo, $t_{\rm coll}$, is the maximum
between the free-fall time,
\begin{equation}
t_{\rm dyn}(r) = [ 3 \pi /32 G \rho(r) ]^{1/2} \ , \label{tdyn}
\end{equation}
and the cooling time
\begin{equation}
t_{\rm cool}(r) = \frac{3}{2} \frac{\rho_{\rm gas}(r)}{\mu m_{p}}
\frac{k \, T}{C n^2_{e}(r) \Lambda(T)}\ , \label{tcool}
\end{equation}
both computed at the virial time $t_{\rm vir}$. In the above
equations $\rho$ is the total matter density, $\rho_{\rm gas}$ is
the gas density, $n_{e}$ is the electron density, $\Lambda(T)$ is
the cooling function and \new{$C = \langle n_e^2(r)
\rangle/\langle n_e(r) \rangle^2$ is the clumping factor, assumed
constant}. In the following we adopt the cooling function given by
Sutherland \& Dopita (1993), which includes the dependence on
metal abundance.

We consider a single-zone galaxy with three gas phases: diffuse
gas in the outer regions, with mass $M_{\rm inf}(t)$, infalling on
a dynamical timescale, cool gas with mass $M_{\rm cold}(t)$,
available to form stars, and hot gas with mass $M_{\rm hot}(t)$,
eventually outflowing. At the virialization we assume that $M_{\rm
inf}(t_{\rm vir})= f_b M_{\rm vir}$, with $f_b=0.16$, the
universal ratio of baryons to DM (Bennett et al. 2003; Spergel et
al. 2003). The infalling gas, initially at the equilibrium
temperature in the DM potential well [Eq.~(\ref{Tvir})], is
transferred to the cool star-forming phase at a rate:
\begin{equation}
\dot M_{\rm cold}(t)=M_{\rm inf}(t)/ {\mathrm max}[t_{\rm
cool}(r_{vir}),t_{\rm dyn}(r_{\rm vir})]. \label{Mdotcold}
\end{equation}
The star formation rate (SFR) is given by:
\begin{equation}
\psi(t) = \int_0 ^{r_{\rm vir}} \frac{1}{{\mathrm max}[t_{\rm
cool}(r),t_{\rm dyn}(r)]} \frac {{\mathrm d}M_{\rm cold}(r,
t)}{{\mathrm d}r} {\mathrm d}r,
\end{equation}
where we assume that the cold gas distribution still follows the
DM distribution. \new{This assumption is admittedly quite
unrealistic since cold cloudlets should rather fall towards the
center. But again, a realistic modelling of the cold gas
distribution would be too ambitious at the present stage.

As mentioned above, according to recent studies (Wechsler et al.
2002, van den Bosch 2002, Zhao et al. 2003) while the build-up of
the potential well is rather fast, the mass of the halo goes on
increasing over a time scale of the order of the Hubble time. The
baryons associated with this slow accretion occupy a large volume
so that have a relatively low density and feel also the heating
effect of the supernova and quasar feedback. It is therefore
reasonable to assume that their cooling time is comparable to, or
longer than, the Hubble time, and does not partake in the star
formation process.

Strictly speaking, in Eq.~(\ref{Mdotcold}) the increase of the
cooling time due to the dilution of the hot gas as the cold gas
drops out should be taken into account. The effect is, however,
minor because the fraction of gas in the cold phase is always $<
50\%$ and can therefore be ignored given the exploratory nature of
the present model.}

Although the Initial Mass Function (IMF) may depend on the
physical properties of the gas, such as density and metallicity
(see Eisenhauer 2001 and references therein), it is often assumed
to be independent of time and galaxy mass. We will consider in the
following the IMF preferred by Romano et al (2002) on the basis of
chemical abundances in local ellipticals, i.e.\ $\Phi(M)$
$\propto$ $M^{-0.4}$ for $M$ $\le$ 1 $M_\odot$ and $\Phi(M)$
$\propto$ $M^{-1.25}$ for $1 \, M_\odot < M \le 100 \, M_\odot$. A
detailed discussion of the effects of varying the IMF on the
properties of the present day elliptical galaxies will be
presented in a forthcoming paper.

The feedback due to supernova (SN) explosions  moves the gas  from
the cold to the hot phase at a rate:
\begin{equation}
\dot M_{\rm cold}^{\rm SN} = -\frac{2}{3} \psi(t) \, \epsilon_{\rm
SN} \,
 \, \frac{\eta_{\rm SN} E_{\rm SN}}{\sigma^2}, \label{eq:snfb}
\end{equation}
where $\eta_{\rm SN}$ is the number of Type II SNe expected per
solar mass of formed stars (determined by the IMF, adopting a
minimum progenitor mass of $8\,M_\odot$), $E_{\rm SN}$ is the
kinetic energy of the ejecta from each supernova ($10^{51}\,$erg;
e.g. Woosley \& Weaver 1986), and $\epsilon_{\rm SN}$ is the
fraction of this energy which is used to reheat the cold gas.
Analyses show that about 90\% of the SN kinetic energy may be lost
by radiative cooling (Thornton et al. 1998; Heckman et al. 2000);
we adopt $\epsilon_{\rm SN}=0.05$ as our reference value,
\new{consistent with the results by  Mac Low \& Ferrara (1999) and
Wada \& Venkatesan (2003)}.

{\new We relate $V_{\rm vir}$ to the line-of-sight velocity
dispersion $\sigma$ using} the relationship $\sigma/V_{\rm
vir}\simeq 0.65$, derived by Ferrarese (2002) for a sample of
local galaxies.

The chemical evolution of the gas is followed by using classical
equations and stellar nucleosynthesis prescriptions, as for
instance reported in Romano et al. (2002).

We can elucidate the dependence of the star formation rate on halo
mass by means of a simple order of magnitude argument. Since the
mean density within the virial radius is $\simeq 200 \rho_{\rm
crit}$, the mean value of $t_{\rm dyn}$ is about a factor of 10
shorter than the expansion timescale, at all redshifts,
independently of the halo mass. If the gas is in virial
equilibrium in the DM potential well, the effective cooling time
of a pure hydrogen plasma is, assuming uniform density:
\begin{eqnarray}
t_{\rm cool, eff} &\!\!\!\! \!\! \simeq &\!\!\!\! \!\! 1.6\,
10^{11} (1+z_{\rm vir})^{-5/2}
h^{1/3} \cdot \nonumber \\
& & \!\!\!\!\!\!\!\!\!\! \!\!\!\!\!\!\!\!\!\!\!\!
\!\!\left({M_{\rm vir}\over
10^{12}M_\odot}\right)^{1/3}\left({M_{\rm vir}/M_{\rm gas}\over
1/0.16}\right)^{1/2}C^{-1}\ \hbox{yr} . \label{tcoolH}
\end{eqnarray}
As discussed by Romano et al. (2002, cf. their Fig.~10), for
spheroidal galaxies with $\lesssim 10^{11}\,M_\odot$ in stars, the
ratio between $M_{\rm vir}$ and the mass in stars at z=0
(including remnants) $M_{\rm sph}$ decreases with increasing
$M_{\rm sph}$, roughly as:
\begin{equation}
M_{\rm vir}/M_{\rm sph} \propto M_{\rm sph}^{-1/3} \ ,
\label{romano}
\end{equation}
due to the larger effect of stellar feedback in shallower
potential wells. Since the mass of cooling gas is approximately
equal to $M_{\rm sph}$, $t_{\rm cool, eff}$ turns out to be very
weakly dependent on $M_{\rm vir}$. For an effective clumping
factor $C_{\rm eff} \gtrsim 10$, Eq.~(\ref{tcoolH}), yields a
cooling time shorter than the mean free-fall time for $z \gtrsim
3$, for all galactic masses. Allowing for the gas metallicity
obviously decreases $t_{\rm cool}$.

\new{If the final mass in stars, $M_{\rm sph}$, is proportional to
the mass of the cooling gas, $M_{\rm sph}$, Eq.~(\ref{romano})
implies that, for galaxies with $M_{\rm sph} < 10^{11}\,M_\odot$,
$M_{\rm vir} \propto M_{\rm sph}^{2/3} \propto M_{\rm gas}^{2/3}$.
Thus} the star formation rate, $\psi(t)\simeq M_{\rm
gas}/\max(t_{\rm cool},t_{\rm dyn})$ is approximately $\propto
M_{\rm vir}^{3/2}$, since both timescales are effectively
independent of, or very weakly dependent on, $M_{\rm vir}$. This
means that stars form faster within larger dark-matter halos, as
in the {\it Anti-hierarchical Baryon Collapse} scenario by Granato
et al. (2001). If, as argued above, the star formation rate is
controlled by $t_{\rm dyn}$, we have $\psi(t)\simeq 32 (M_{\rm
gas}/ 10^{11} M_\odot)(1+z)^{3/2}\,M_\odot\,\hbox{yr}^{-1}$.

\subsection{Black-hole growth}

As discussed by Haiman et al. (2003), the available information on
the evolution of both the global star formation rate and the
quasar emissivity is broadly consistent with the hypothesis that
star formation in spheroids and black-hole (BH) fuelling are
proportional to one another.

One mechanism yielding such proportionality has been discussed by
Umemura (2001), Kawakatu \& Umemura (2002) and Kawakatu et al.
(2003). In the central regions of proto-galaxies the drag due to
stellar radiation may result in a loss of angular momentum of the
gas at a rate that in a clumpy medium is well approximated by
\begin{equation}
\frac{d ln J}{dt}\simeq \frac{L_{\rm sph}} {c^2 M_{\rm
gas}}(1-e^{-\tau})\ , \label{dJdt}
\end{equation}
where $L_{\rm sph}$ is the global stellar luminosity and $\tau$ is
the effective optical depth of the spheroid. The latter quantity
is given by $\tau=\bar{\tau} N_{\rm int}$, where $\bar{\tau}$ is
the average optical depth of single clouds and $N_{\rm int}$ is
the average number of clouds intersected by a light ray over a
typical galactic path.

The gas can then flow towards the center, feeding a mass reservoir
around the BH at a rate (Kawakatu et al. 2003)
\begin{equation}
\dot M_{\rm inflow} \simeq - M_{\rm gas}\frac {d ln J}{dt}\simeq
\left(\frac{L_{\rm sph}} {c^2}\right)(1-e^{-\tau})\ .
\label{dotMres}
\end{equation}
During the early evolutionary stages the luminosity is dominated
by massive main sequence stars, $M\geq 5 M_{\odot}$, and is thus
proportional to the star formation rate $\psi(t)$. For the adopted
IMF we have:
\begin{equation}
\dot M_{\rm inflow} \simeq 1.2 \times 10^{-3}\psi (t)(1-e^{-\tau})
\ M_{\odot}\,\hbox{yr}^{-1}\ . \label{inflowrate}
\end{equation}
While this expression is useful for the analytical estimates
presented in Sect.~3.1, in the full calculations we have adopted
the values $L_{\rm sph}$ computed using our spectrophotometric
code GRASIL. We parameterize the optical depth $\tau$ as:
\begin{equation}
\tau = \tau_{0} \left(\frac {Z}{Z_{\odot}}\right) \left(\frac
{M_{\rm gas}}{10^{12}M_\odot}\right)^{\frac{1}{3}}, \label{tau}
\end{equation}
and for $\tau_0$ we explore the range from 1 to 10 (see Sect.~3).

An order of magnitude estimate of the relevant timescales and
luminosities can be derived assuming that $\dot M_{\rm inflow}$ is
directly the accretion rate on the central BH. Then, the timescale
for it to grow to a mass $M_{\rm BH}$ is, using the order of
magnitude estimate of $\psi (t)$ given in the previous section,
\begin{eqnarray}
t_{\rm BH} &\simeq & {M_{\rm BH} \over \dot M_{\rm inflow}} \simeq
2.3 \times 10^9 {M_{\rm BH}\over 10^8 M_\odot}\left({M_{\rm
gas}\over 10^{11}\, M_\odot}\right)^{-1} \cdot \nonumber \\
&\cdot& (1+z)^{-3/2} \left(1-e^{-\tau}\right)^{-1} \mbox{yr}\ .
\label{tBH}
\end{eqnarray}
Thus, if the accretion is not limited e.g. by radiation pressure
or by angular momentum, black-holes can grow to very large masses
from small (e.g., stellar mass) seeds in a time shorter than the
age of the universe at all relevant redshifts. If $M_{\rm
BH}\propto M_{\rm gas}$, the growth time is independent of $M_{\rm
BH}$.

For accretion with radiative efficiency $\eta$, the bolometric
luminosity is $L_{\rm bol}\simeq \eta \dot M_{\rm inflow} c^2$.
\new{Following Elvis, Risaliti, \& Zamorani (2002), we adopt, as a
reference value, $\eta=0.15$}. In units of the Eddington
luminosity
\begin{equation}
L_{\rm Edd} \simeq 1.26\times 10^{46} (M_{\rm BH}(t)/10^8
M_\odot)\,\hbox{erg}\,\hbox{s}^{-1},
\end{equation}
and setting $M_{\rm BH}(t)\simeq \dot M_{\rm inflow} t$, we have:
\begin{equation}
{L_{\rm bol} \over L_{\rm Edd}} \simeq 3 {\eta \over 0.15}
\left({t \over 10^8\,\hbox{yr}}\right)^{-1} \ , \label{Eddratio}
\end{equation}
showing that the fast accretion phase and the build up of
super-massive BH can be quite fast, thus accounting for QSO at
high z.

Obviously, super-Eddington accretion is not the only mechanisms
for a rapid black-hole growth. At the other extreme we may have
accretion with low radiative efficiency such as the black-hole
mergers at high redshifts advocated by Haiman et al. (2003),
\new{which may be testable by gravitational wave experiments
like LISA (Menou, Haiman \& Narayan 2001).} Since these early
evolutionary phases are expected to be heavily dust obscured, a
powerful tool to discriminate among the various possibilities is
hard X-ray emission. Far-IR/mm observations are also useful, but
less capable of distinguishing the effect of the AGN from that of
a starburst.

While mechanisms for super-Eddington accretion have been proposed
(Begelman 2001, 2002), fully unperturbed free-fall is unrealistic.
Correspondingly, $L_{\rm bol}/L_{\rm Edd}$ will not reach the
extreme values indicated by Eq.~(\ref{Eddratio}) and the duration
of the $L_{\rm bol}/L_{\rm Edd}\gtrsim 1$ phase can be longer. In
our model, we first let the material infalling at the rate given
by Eq.~(\ref{dotMres}) accumulate in a circum-nuclear mass
reservoir, and then flow toward the black hole on a time scale
depending on the viscous drag $\tau_{\rm visc}\sim r^2/\nu$.
\new{The reservoir accumulates mass at a net rate $\dot M_{\rm res}=
\dot M_{\rm inflow}-\dot M_{\rm BH}$, where $\dot M_{\rm BH}$ is
computed as follows}. Following Duschl et al. (2000) and Burkert
\& Silk (2001) we adopt a viscosity $\nu={\cal R}^{-1}_{\rm crit}
v_r$, where ${\cal R}_{\rm crit}=100$--1000 is the critical
Reynolds number for the onset of turbulence. With these
assumptions the viscous time can be expressed as $\tau _{\rm
visc}= \tau _{\rm dyn} {\cal R}_{\rm crit}$. The dynamical time
$\tau _{\rm dyn}=({3 \pi}/{32 G \rho_s})^{{1}/{2}}$ is referred to
the system `black hole plus reservoir'.

The accretion radius of the BH is given by $r_a={G M_{BH}}/{V_{\rm
vir} ^2}$. Defining the reservoir dimension $R_{\rm res}=\alpha
r_a$, we \new{estimate $\rho_s$ as the mean density within a
sphere of radius $R_{\rm res}$ with mass $M_{\rm BH}+M_{\rm res}$,
to} get:
\begin{equation}
\tau _{\rm dyn}= \pi \left({\alpha\over 2}\right) ^{3/2} \frac
{G}{V_{\rm vir}^3} \frac {M_{\rm BH}^{3/2}} {(M_{\rm BH}+M_{\rm
res})^{1/2}}.
\end{equation}
\new{ The viscous accretion rate onto the BH can be defined as
\begin{equation}
\dot M_{\rm BH}^{\rm visc}= \frac {M_{\rm res}}{\tau _{\rm visc}}=
k_{\rm accr} \frac {\sigma ^3}{G} \left(\frac {M_{\rm res}}{M_{\rm
BH}}\right)^{3/2} \left(1+\frac {M_{\rm BH}}{M_{\rm
res}}\right)^{1/2}. \label{eq:mdotaccvis}
\end{equation}
The constant $k_{\rm accr}=[\pi (\alpha/2)^{3/2}(V_{\rm
vir}/\sigma)^3 {\cal R}_{\rm crit}]^{-1}$ has a rather wide range
of possible values. For ${\cal R}_{\rm crit}=100$ and $\alpha=10$
we have $k_{\rm accr}\sim 10^{-4}$ that we adopt as the reference
value in the following. The actual accretion rate is then given by
\begin{equation}
\dot M_{\rm BH}=\min{(\dot M_{\rm BH}^{visc}, {\cal A} M_{Edd})}
\label{eq:mdotacc}
\end{equation}
where $M_{Edd}=L_{Edd}/(\eta c^2)$ and  ${\cal A}=(L/L_{\rm
Edd})_{\rm max}$ is the maximum allowed Eddington ratio. In our
computations, we allow at most mildly super-Eddington accretion,
i.e.\ ${\cal A}\le \mbox{a few}$. }

\subsection{QSO feedback}

The QSO activity affects the interstellar medium of the host
galaxy and also the surrounding intergalactic medium through both
the radiative output and the injection of kinetic energy producing
powerful gas outflows. Quasar-driven outflows have been invoked to
produce an intergalactic magnetic field (Furlanetto \& Loeb 2001),
and to preheat the intra-cluster medium (Valageas \& Silk 1999;
Kravtsov \& Yepes 2000; Wu et al. 2000; Bower et al. 2001;
Cavaliere et al. 2002; Platania et al. 2002). In the case of radio
loud QSOs there is evidence that up to about half of the total
power is in the jets (Rawlings \& Saunders 1991; Celotti et al.
1997; Tavecchio et al. 2000). In broad absorption line (BAL) QSOs,
the kinetic power of the outflowing gas can be a significant
fraction of the bolometric luminosity (Begelman 2003). X-ray
observations of BAL QSOs revealed significant absorption
($N_H\gtrsim 10^{23}\,\hbox{cm}^{-2}$), implying large outflows
($\dot M_{\rm out}\gtrsim 5 \ M_{\odot}/$yr) and large kinetic
luminosities $L_K$ (Brandt et al. 2001; Brandt \& Gallagher 2000).

Theoretical studies on the mechanisms responsible for AGN-driven
outflows show that efficient acceleration could be due to
radiation pressure through scattering and absorption by dust (see
e.g. Voit et al. 1993) and scattering in resonance lines (see e.g.
Arav, Li \& Begelman 1994). Murray et al. (1995) presented a
dynamical model for a wind produced just over the disk by a
combination of radiation and gas pressure. In a similar way Proga,
Stone \& Kallman (2000) showed that a wind can be launched from a
disc around a $10^8\, M_{\odot}$ black hole with velocities up to
0.1c and mass-loss rate of $0.5\, M_{\odot}\, \hbox{yr}^{-1}$.
Following Murray et al. (1995), an approximate solution for the
wind velocity produced by line acceleration as a function of the
radius is:
\begin{equation}
v=v_{\infty}\left(1-\frac {r_f} {r}\right)^{2.35}
\end{equation}
where $r_f$ is the radius at which the wind is launched.

The asymptotic speed is
\begin{equation}
v_{\infty}\sim \left (\gamma \frac{GM_{\rm BH}}{r_f}\right)^{1/2}
\end{equation}
where $\gamma$ is  related to the force multiplier (see e.g. Laor
\& Brandt 2002). Adopting the reference values of the parameters
of Murray et al. (1995), we have $\gamma\simeq 3.5$. By replacing
the BH mass with the corresponding Eddington luminosity $L_{{\rm
Edd},46}$ in units of $10^{46}$ $\hbox{erg}\,\hbox{s}^{-1}$,  we
get:
\begin{equation}
{v_{\infty}\over c}\sim 6.2 \times 10^{-2} \left(\frac
{r_f}{10^{16}\,\hbox{cm}}\right)^{-1/2} {L_{{\rm Edd},46}}^{1/2} .
\end{equation}
Detection of outflows with velocities ranging between 0.1c and
0.4c in the Broad Absorption Line (BAL) quasars APM$08279+0522$
(Chartas et al. 2002) and PG$1115+080$ (Chartas et al. 2003) has
been reported, based on observations of X-ray broad absorption
lines performed with the Chandra and XMM-Newton X-ray
observatories. In the case of APM$08279+0522$ an intrinsic
bolometric luminosity of $L_{\rm bol}\sim 2.4\times
10^{47}\,\hbox{erg}\, \hbox{s}^{-1}$ has been estimated by Egami
et al. (2000). Under the assumption of $L_{\rm Edd} \sim 1$--$3\
L_{\rm bol}$, the maximum observed velocity $v\simeq 0.4c$ is
obtained with a launching radius $r_f\sim 0.5$--$2 \times
10^{16}\,\hbox{cm}$. Such small values of $r_f$ are confirmed by
the observed variability of the absorption line energies and
widths in APM$08279+0522$ over a proper timescale of 1.8 weeks
(Chartas et al. 2003).

The asymptotic speed is reached at $r\gtrsim 40 r_f$. If $f_c$ is
the covering factor of the outflow and using $\dot M_w= 4\pi r^2
\rho(r) v_r \sim 4 \pi f_c m_H N_H 40 r_f v_{\infty}$, we get
\begin{equation}
\dot M_w=2.6 f_c N_{22} L_{{\rm Edd},46}^{1/2}
\left(\frac{r_f}{10^{16}\,\hbox{cm}}\right)^{1/2}\ \
M_{\odot}\,\hbox{yr}^{-1}
\end{equation}
where and $N_{22}= {N_H}/{10^{22}\,\hbox{cm}^{-2}}$. Adopting
$r_f=1.5\times 10^{16}$ cm as a reference value, the kinetic power
in the outflow is
\begin{equation}
L_K=\frac {1}{2} \dot M_{w} v_{\infty}^2\simeq 3.6 \times 10^{44}
f_c N_{22} L_{{\rm Edd},46}^{3/2}\,\hbox{erg}\,\hbox{s}^{-1}\, ,
\label{LK}
\end{equation}
and may thus amount to several percent of the accretion luminosity
for highly luminous QSOs. Interestingly, this corresponds to the
power required to account for the pre-heating of the intra-cluster
medium (Bower et al. 2001; Platania et al. 2002; Lapi et al.
2003). It is also interesting to notice that ${\dot M_{w}}/{\dot
M_{\rm acc}} \simeq 2.5 ({\epsilon}/{0.1}) f_c N_{22} L_{{\rm
Edd},46}^{-1/2}$, implying that for highly luminous QSOs $\dot M_w
\sim \dot M_{\rm acc} $, if $f_c N_{22}\sim 1$. High luminosity
QSOs emitting at the Eddington limit  are able to generate winds
involving relatively small amounts of gas, but with very high
velocities and significant kinetic energies. When the luminosity
decreases below the Eddington limit we replace  the Eddington
luminosity with the bolometric luminosity in Eqs.\ (\ref{LK}) and
(\ref{eq:mdotfbqso}).

Estimating the fraction of the kinetic luminosity $L_K$
transferred to the interstellar medium is a rather complex
problem. One possible effect of outflows and jets is to transport
the ambient gas to larger radii. This effect has been recently
estimated through numerical simulations by Br\"uggen et al.
(2002). By investigating the effects of radio cocoons on the
intra-cluster medium, they concluded that frequent low-power
activity cycles are rather efficient in stirring the environment
gas, particularly in regions close to the injection point of the
high speed gas. En{\ss}lin \& Kaiser (2000) evaluated the energy
accumulated in the cocoons around radio-galaxies. Inoue \& Sasaki
(2001) argued that for non relativistic cocoon plasma, the
fraction of the jet energy deposited in the intra-cluster medium
can reach $40\%$, neglecting the radiative cooling. Similar
conclusions have been reached by Bicknell et al. (1997). Nath \&
Roychowdhury (2002), taking into account also the radiative
losses, found that a large fraction ($\gtrsim 0.5$) of the kinetic
energy  of BAL and radio loud QSOs is transferred to an ambient
gas with number density $n\simeq 0.1$--$1\,\hbox{cm}^{-3}$ and
temperature $T_{\rm vir}\simeq 10^6$ K. These values of density
and temperature are quite similar to those of the gas in the outer
regions of massive galactic halos.

We assume that the QSO feedback heats up the interstellar medium
at a rate $L_h=f_hL_K$ and removes it from the cold phase at a
rate
\begin{equation}
\dot M_{\rm cold}^{\rm QSO}=-\frac{2}{3} \frac {L_h}{\sigma^2}
\frac{M_{\rm cold}} {M_{\rm gas}}\ , \label{ratehot}
\end{equation}
where $M_{\rm gas}=M_{\rm cold}+M_{\rm inf}$ is the mass of the
gas in the cold phase plus that of the gas which has not yet
fallen in the star forming region. Setting $\epsilon _{\rm
QSO}=({f_h}/{0.5}) ({f_c}/{0.1}) ({N_{22}}/{10})$, we get
\begin{equation}
\dot M_{\rm cold}^{\rm QSO}\simeq -2 \times 10^{3} \frac
{\epsilon_{\rm QSO} L_{{\rm Edd}, 46}^{3/2}}{(\sigma/300\,
\hbox{km}\,\hbox{s}^{-1})^2}\frac{M_{\rm cold}} {M_{\rm gas}} \ \
M_{\odot}\,\hbox{yr}^{-1} \ , \label{eq:mdotfbqso}
\end{equation}
We have explored the range $1\leq \epsilon_{\rm QSO}\leq 10$,
adopting $\epsilon_{\rm QSO}=6$ as our reference value. The
fraction ${M_{\rm cold}}/{M_{\rm gas}}$ in the right-hand side of
Eq.~(\ref{ratehot}) has been introduced in order to share the QSO
feedback between the cold and the infalling gas. Thus the QSO
feedback also removes the infalling gas at a rate
\begin{equation}
\dot M_{\rm inf}^{\rm QSO}=-\frac{2}{3} \frac {L_h}{\sigma^2}
\frac{M_{\rm inf}} {M_{\rm gas}}\ .
\end{equation}
\new{The quasar feedback can easily heat the interstellar gas to
temperatures $\sim 1\,$keV (Valageas \& Silk 1999; Bower et al.
2001; Nath \& Roychowdury 2002; Platania et al. 2002; Lapi et al.
2003), thus unbinding it and making it flow into the intergalactic
medium (IGM). The corresponding energy injection in the IGM may
account for the steepening of the X-ray luminosity--temperature
correlation observed in groups of galaxies (O'Sullivan, Ponman, \&
Collins 2003). Given the low IGM gas density, only a small
fraction of it will eventually cool down and fall back again to
form stars.  }

\begin{figure}[t]
\epsscale{0.9} \plotone{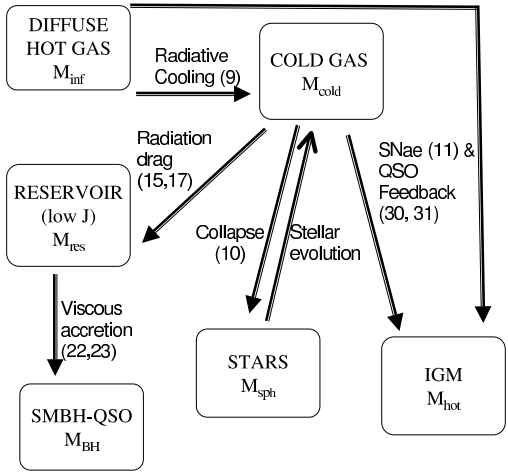} \caption{Scheme of the baryonic
components included in the model (boxes), and of the corresponding
mass transfer processes (arrows). \new{The numbers near the arrows
point to the main equations describing those processes.}}
\label{schema}
\end{figure}

\begin{table}[!htb] 
\begin{center}
\caption{Model parameters, reference value and number of the
equation where it is introduced (or short description).
 \label{table:galfor}}
\begin{tabular}{lcc}
\tableline
\multicolumn{1}{c}{Symbol} & \multicolumn{1}{c}{Value} & \multicolumn{1}{c}{Eq.}\\
\tableline
$C$ & 20 & \ref{tcool}\\
$\epsilon_{\rm SN}$ & 0.05 & \ref{eq:snfb}\\
$\tau_0$ & 4 & \ref{tau}\\
$k_{\rm accr}$ & $1\times 10^{-4}$ & \ref{eq:mdotaccvis}\\
$M_{\rm BHseed}$ & $10^3\,M_\odot$ & Seed BH mass\\
$(L/L_{\rm Edd})_{\rm max}$ & 3 & \ref{eq:mdotacc}\\
$\epsilon_{\rm QSO}$ & $6.0$ & \ref{eq:mdotfbqso}\\
$\eta$ & $0.15$ & \ref{Eddratio}\\
\tableline
\end{tabular}
\end{center}
\end{table}

\section{Results}

\new{The results presented here refer to halos with mass in the
range $2.5 \times 10^{11}\, M_\odot\lesssim M_{\rm vir}\lesssim
1.6 \times 10^{13}\,M_\odot$, formed at $z_{\rm vir} \gtrsim 1.5$.
This lower limit to the virialization redshift allows us to
crudely filter out halos hosting discs and irregular galaxies. In
our view, late-virialized objects had more time to acquire a
substantial angular momentum and are less likely to take on an
early-type morphology. This is in keeping with the common notion
that stellar populations in disks are on average substantially
younger than those in spheroids.}

The model, whose main baryonic components and corresponding mass
transfer are sketched in Fig.~\ref{schema}, allows us to follow
the star formation and chemical enrichment histories of spheroidal
galaxies, as well as the growth of their central BHs, once the
virialization time and the halo mass is given. Interfacing it with
the code GRASIL (see Silva et al.\ 1998 for details), which
computes the spectrophotometric evolution from radio to X-ray
wavelengths including the dust effects, we get the spectral
properties of the spheroidal galaxies as a function of cosmic
time. In the application of GRASIL, we have made the standard
assumption that the starburst is highly obscured throughout its
duration, with an 1 $\mu$m optical depth of molecular clouds,
where new stars are born, of $\tau_{1} =30$ for solar metallicity,
and scaling linearly with $Z$. As for the dust emissivity index,
we have adopted the canonical value of 2. The results presented in
this paper are only weakly sensitive to variations of $\tau_{1}$
by a factor of several.

The relationship between $\tau_{1}$ and the optical depth given by
Eq.~(\ref{tau}) is not straightforward. The starlight is processed
by dust within molecular clouds and re-emitted in the far-IR, with
a peak at 60--$100\,\mu$m. At these wavelengths, the optical depth
is lower than that at $1\,\mu$m by a factor $\sim 10^{-2}$. The
effective optical depth to the inter-cloud radiation, responsible
for the drag exerted on gas clouds as discussed in Sect.~2.3, is
therefore $\tau \sim 10^{-2} \tau_1 N_{\rm int}$, where $N_{\rm
int}$ is the mean number of clouds intersected by a photon
(Kawakatu \& Umemura 2002). The latter number depends, among other
things, on the radial distribution of the clouds within the
galaxy. For instance, a population of giant molecular clouds with
a typical size of $10\,$pc, masses of the order of $10^5
M_{\sun}$, uniformly distributed in the central $\sim 10$ kpc of
the galaxy, give $N_{\rm int} \sim 10$, for $M_{\rm gas} \simeq
10^{12}\,M_\odot$. As anticipated, for $\tau_0$ [Eq.~(\ref{tau})]
we explored the range 1--10.

In this Section we present some of the most important results,
obtained using the reference values of the parameters already
discussed and \new{summarized in Table \ref{table:galfor}}. The
discussion of the effects of varying the parameters within the
allowed ranges will be presented in the next Section.

\begin{figure}[t]
\epsscale{0.6} \plotone{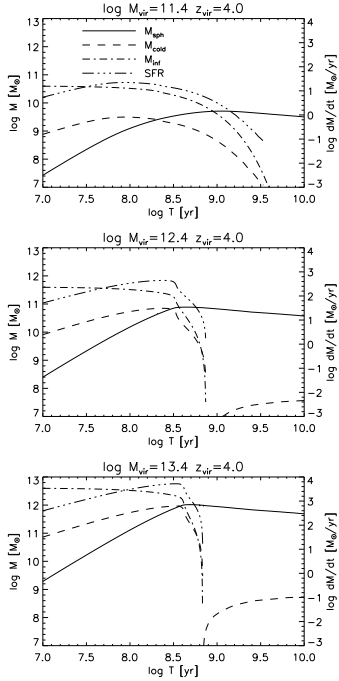} \caption{Evolution of the mass in
stars, in the cold ($M_{\rm cold}$) and infalling ($M_{\rm inf}$)
gas components, and of the Star Formation Rate, as a function of
the galactic age $T$, for 3 values of $M_{\rm vir}$ and $z_{\rm
vir} =4$. Note the sharp break corresponding to the sweeping out
of the interstellar medium by the quasar feedback, for large
$M_{\rm vir}$. } \label{sfrlt}
\end{figure}

\begin{figure}[t]
\epsscale{0.6} \plotone{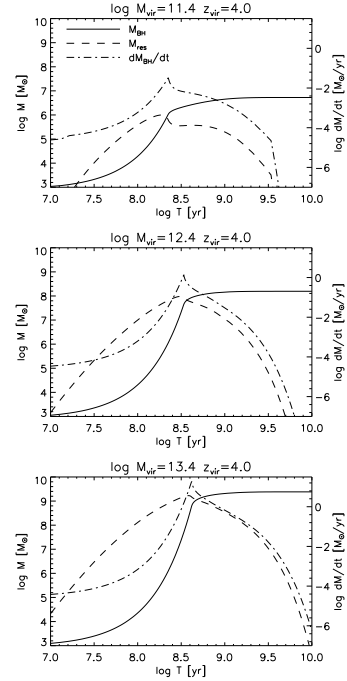} \caption{Growth of the black-hole
and of the reservoir as a function of the galactic age $T$, for
the same values of $M_{\rm vir}$ and $z_{\rm vir} =4$ as in
Fig.~\ref{sfrlt}. The apex of $dM_{\rm BH}/dt$ corresponds to the
end of the exponential growth of the black-hole mass, when the
mass accretion rate becomes insufficient to support the adopted
maximum Eddington ratio, $(L/L_{\rm Edd})_{\rm max}=3$.}
\label{bhlt}
\end{figure}

Examples of the time evolution of the various components for
different virialization redshifts and halo masses are shown in
Fig.~\ref{sfrlt} and Fig.~\ref{bhlt}.

\subsection{Time delay between star formation and QSO activity}

A key result of the model is the prediction of the time delay
between the onset of vigorous star formation, at the virialization
epoch, and the peak of the QSO activity.  The combined action of
stellar and nuclear feedbacks, eventually sweeping out the
interstellar medium, determines the duration of the star
formation.

\begin{figure}[t]
\epsscale{1.} \plotone{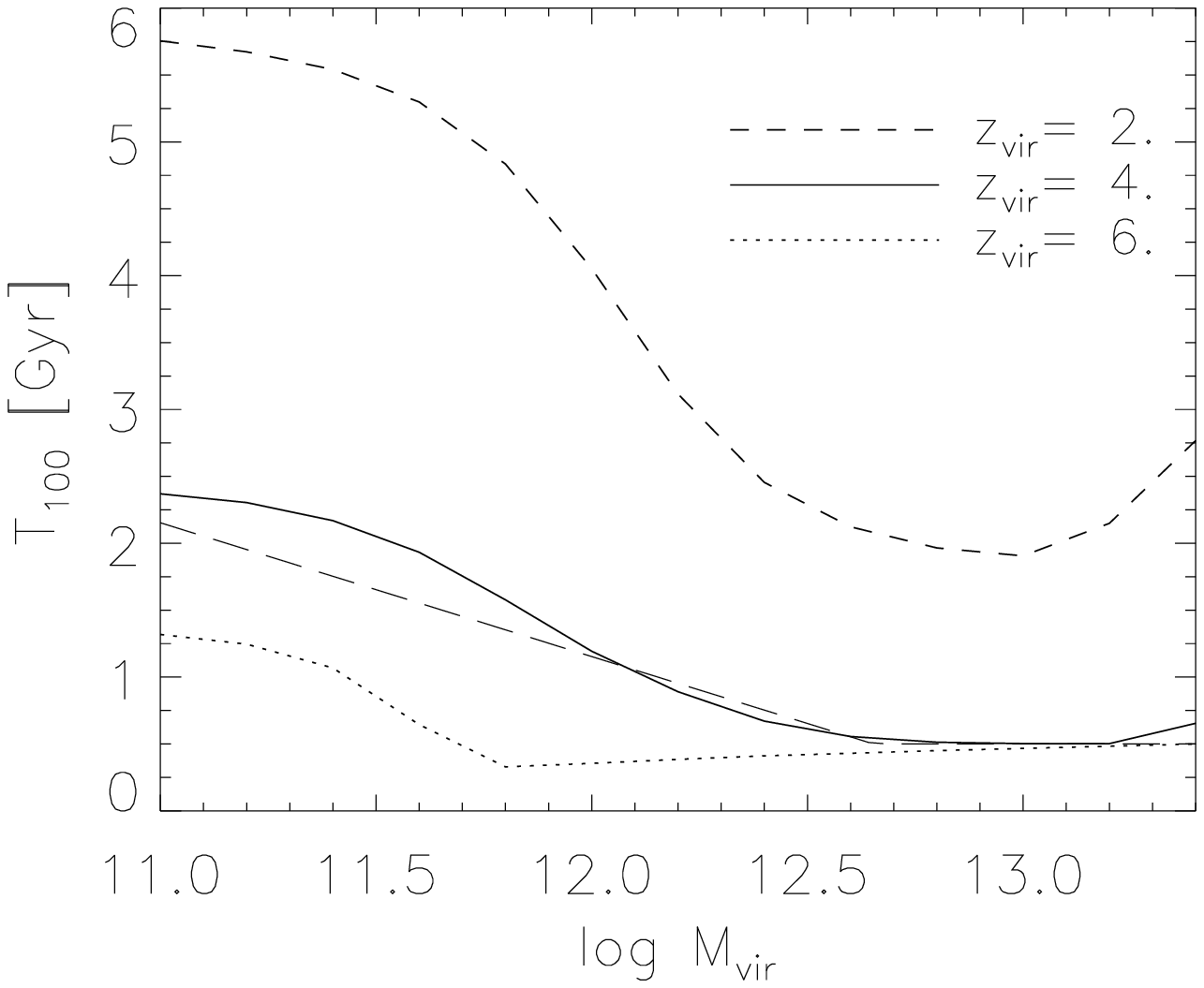} \caption{Duration of the
star-formation burst as a function of $M_{\rm vir}$, for 3 values
of $z_{\rm vir}$ (see text for the exact definition of $T_{100}$).
\new{For our choice of the parameters, at low $z_{\rm vir}$ the
cooling time becomes longer than the dynamical time at large
masses, causing the upturn of $T_{100}$}.  The long-dashed line
close to the solid line corresponding to $z_{\rm vir}=4$
represents the empirical recipe adopted by Granato et al. (2001).
} \label{T100}
\end{figure}

\begin{figure}[t]
\epsscale{1.} \plotone{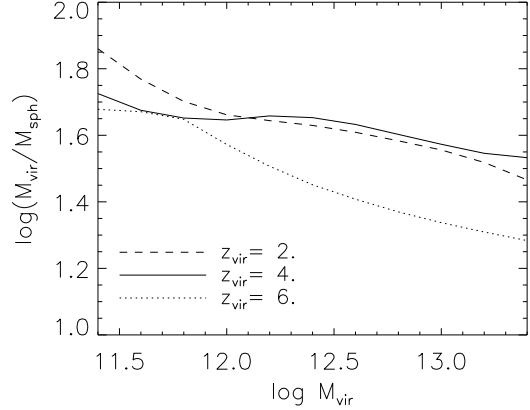} \caption{Virial to stellar mass
ratio at the present time as a function of $M_{\rm vir}$ for 3
values of $z_{\rm vir}$. } \label{MdmsuMsph}
\end{figure}

As shown in Fig.~\ref{T100} the duration of the most active star
formation  phase decreases with increasing $M_{\rm vir}$ and
$z_{\rm vir}$. The timescale $T_{100}$ is defined as the galactic
age at which the gas mass still available for star formation or
accretion, i.e.\ $M_{\rm cool} + M_{\rm inf}$, is reduced to $1\%$
of the initial value. This corresponds approximately to the
duration of the star formation burst empirically estimated by
Granato et al. (2001), and given by their Eq.~(8). In the redshift
range $3\lesssim z_{\rm vir} \lesssim 6$, $T_{100}\simeq
0.5$--$1\,$Gyr for $M_{\rm vir} \gtrsim 10^{12}\, M_{\odot}$,
corresponding (see Fig.~\ref{MdmsuMsph}) to $M_{\rm sph} \gtrsim
2.5 \times 10^{10}\, M_{\odot}$. For smaller masses the duration
of the actively star forming phase increases to reach 1.5-3 Gyr
for $M_{\rm sph} \simeq 3 \times 10^{9}\, M_{\odot}$, if
virialized at $z_{\rm vir} \gtrsim 3$, and is even longer if
virialization occurs at lower $z$.

A feeling of the role of the main ingredients of the model in
shaping this result can be obtained ignoring the reservoir, so
that the inflow onto the central black hole is governed by
Eq.~(\ref{dotMres}). The inflow is definitively halted at a time
$t_{\rm out}$ when the energy injected by the QSO in the
interstellar medium equals the binding energy of the gas (cf.
Cavaliere et al. 2002):
\begin{equation}
\int_0^{t_{\rm out}} f_h L_K dt = M_{\rm vir} V^2_{\rm vir}\ ,
\label{energybalance}
\end{equation}
where $V_{\rm vir}$ is related to the mass inside the virial
radius, $M_{\rm vir}$, by (Bullock et al. 2001b):
\begin{equation}
V_{\rm vir}= 75 (1+z_{\rm vir}) ^{1/2} \left({M_{\rm vir} \over
10^{11} h^{-1} M_\odot}\right)^{1/3} \,\hbox{km}\,\hbox{s}^{-1}\ .
\label{Vvir}
\end{equation}
Using Eq.~(\ref{LK}) and Eq.~(\ref{tBH}) with $\exp(-\tau)\ll 1$,
and assuming that the black hole growth occurs in a time shorter
than the expansion timescale, we obtain:
\begin{eqnarray}
t_{\rm out}&\!\!\!\!\!\! \simeq &\!\!\!\!\!\!  7.9\times 10^8
h^{4/15} f_h^{-2/5}(1+z_{\rm vir})^{-1/2}  \cdot \nonumber \\
&\!\!\!\!\!\! \!\!\!\!\!\! \cdot& \!\!\!\!\!\!  \left({M_{\rm vir}
\over 10^{12} M_\odot}\right)^{2/3}
\left({M_{\rm gas}\over 10^{11}M_\odot}\right)^{-3/5} \nonumber \\
&\simeq & 1.2\times 10^9 h^{4/15} f_h^{-2/5}  \cdot \nonumber \\
&\cdot& (1+z_{\rm vir}) ^{-1/2} \left({M_{\rm vir} \over 10^{12}
M_\odot}\right)^{-7/30} \, \hbox{yr} \ , \label{tout}
\end{eqnarray}
where the last step follows from Eq.~(\ref{romano}), normalized to
$M_{\rm vir}/M_{\rm sph} \simeq 20$ for $M_{\rm sph}=5\times
10^{10}M_\odot$, assuming $M_{\rm gas}\simeq M_{\rm sph}$.

\begin{figure}[t]
\epsscale{1.} \plotone{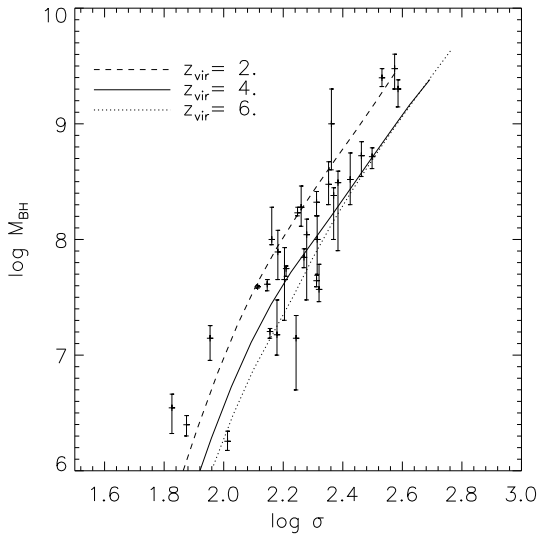} \caption{Predicted relationship
between black-hole mass and line-of-sight velocity dispersion of
the host galaxy for different virialization redshifts. Data are
from Tremaine et al. (2002).} \label{MBHsigma}
\end{figure}

The final black-hole mass (except for the effect of later
re-activation phases) can be roughly estimated to be $\sim M_{\rm
BH}(t_{\rm out})$:
\begin{eqnarray}
M_{\rm BH}&&\!\!\!\!\!\!  \!\!\!\!\!\!\!\!\! (t_{\rm out}) \simeq
4.3 \times 10^7 h^{4/15}
f_h^{-2/5} \cdot \nonumber \\
&\!\!\!\!\!\! \!\!\!\!\!\!\!\!\!\!\!\! \!\!\!\!\!\!
\cdot&\!\!\!\!\!\! \!\!\!\!\!\! \!\!\!\!\!\!\left({M_{\rm vir}
\over 10^{12} M_\odot}\right)^{2/3}\left({M_{\rm
gas}\over 10^{11}M_\odot}\right)^{2/5}(1+z_{\rm vir})\, M_\odot \nonumber \\
\!\!\!\!\!\! \!\!\!\!\!\! \!\!\!\!\!\! \!\!\!\!\!\! &\simeq
&3.2\times 10^7 h^{4/15} f_h^{-2/5}  \cdot \nonumber \\
\!\!\!\!\!\! \!\!\!\!\!\! \!\!\!\!\!\! &\cdot&\!\!\!\!\!\!
\left({M_{\rm vir} \over
10^{12} M_\odot}\right)^{19/15}(1+z_{\rm vir}) \, M_\odot \nonumber \\
\!\!\!\!\!\! \!\!\!\!\!\! \!\!\!\!\!\! \!\!\!\!\!\! &\simeq &
2.5\times 10^8 \left({h\over 0.7}\right)^{-1} \left({f_h\over
0.3}\right)^{-2/5} \cdot \nonumber \\
\!\!\!\!\!\! \!\!\!\!\!\! \!\!\!\!\!\! \!\!\!\!\!\!\!\!\!\!\!\!
&\cdot&\!\!\!\!\!\! \left({\sigma \over
200\hbox{km}/\hbox{s}}\right)^{19/5}\left({1+z_{\rm vir}\over
4}\right)^{-9/10}\!\!\!\!\!\! M_\odot \ , \label{MBHfinal}
\end{eqnarray}
where we have used Eqs.~(\ref{romano}) and (\ref{Vvir}), and have
adopted the relationship $\sigma \simeq 0.65 V_{\rm vir}$
(Ferrarese 2002). This result is in remarkably good agreement with
the determination by Gebhardt et al. (2000) of the $M_{\rm
BH}$--$\sigma$ relationship. A full comparison of our model
predictions with observational data is shown in
Fig.~\ref{MBHsigma}, where the expected spread due to the
distribution of virialization redshifts is also illustrated. Note
the predicted fall-off of $M_{\rm BH}$ at low values of $\sigma$,
due to the combined effect of SN feedback which is increasingly
efficient with decreasing halo mass in slowing down the gas infall
onto the central BH, and of the decreased radiation drag due to a
decrease of $\tau$ [Eq.~(\ref{tau})], which is, for these objects,
$\ll 1$. This steepening of the lower part of the $M_{\rm
BH}$--$\sigma$ relationship translates in a {\it flattening} of
the lower part of the $M_{\rm vir}$--$M_{\rm BH}$ correlation, in
agreement with the results reported in Fig.~5 of Ferrarese (2002).

\begin{figure}
\epsscale{1.} \plotone{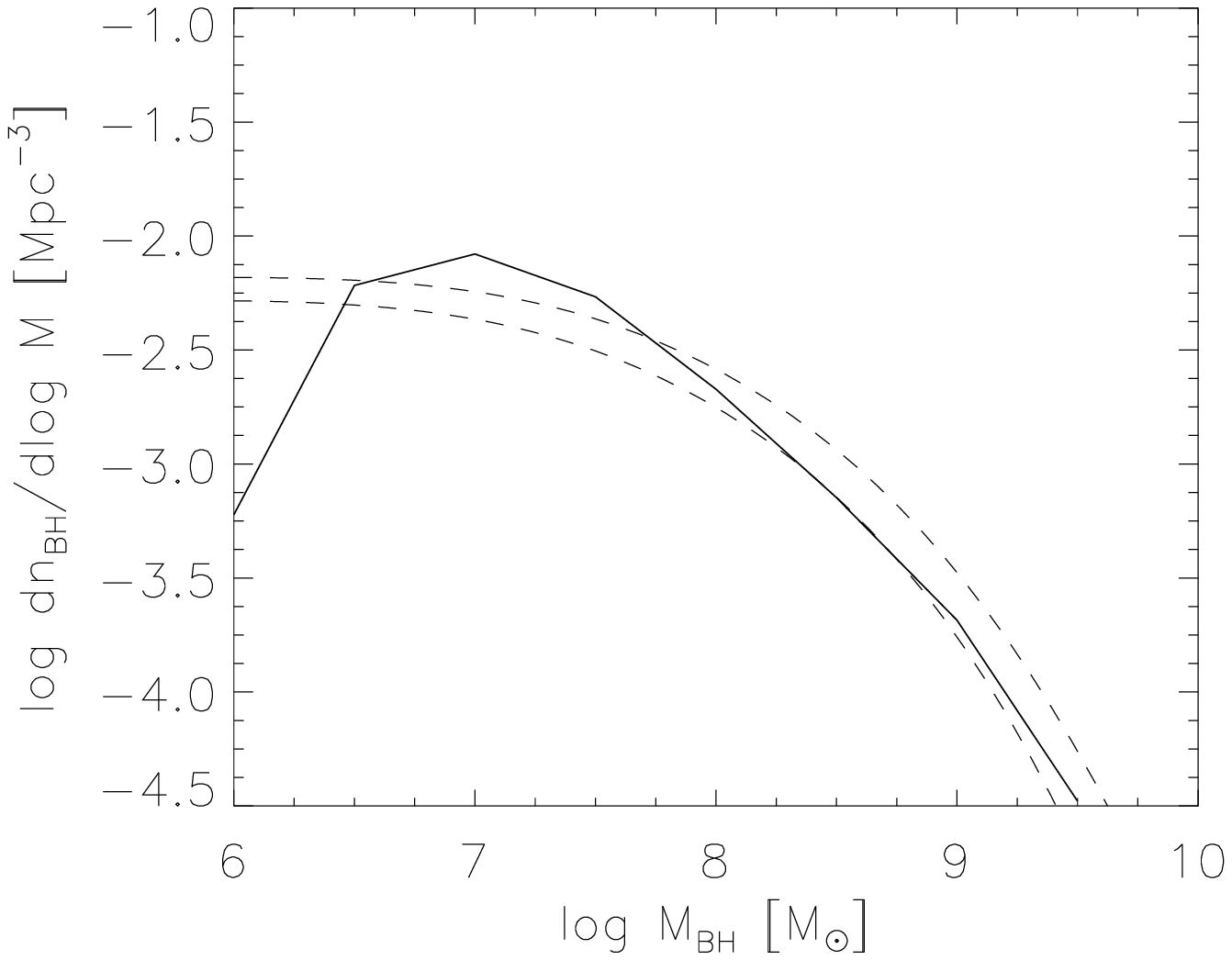} \caption{Predicted local black-hole
mass function (solid line) compared with the recent estimate by
Shankar et al. (in preparation, \new{region delimited by the two
thin dashed lines}). The total mass density in BHs they derive is
$\rho_{BH} \simeq 4 \times 10^5 M_{\odot} \mbox{Mpc}^{-3}$, 25\%
less than our model. The decline of the model at low $M_{\rm BH}$
is due to having considered only objects with $M_{\rm vir} \geq
2.5\times 10^{11} M\odot$. } \label{BHmassfunction}
\end{figure}

In Fig.~\ref{BHmassfunction} the predicted BH mass function at
$z=0$ is shown against the local BH mass function recently derived
by Shankar et al. (in preparation) following Salucci et al.
(1999). The total local BH mass density is estimated to be
$5.2\times 10^5\,M_\odot\,\hbox{Mpc}^{-3}$.

The physics that rules the fast growth of the central BH in
massive halos is the key point: its growth is paralleled by \new{
an increasing feedback from the nuclear activity. In the most
massive halos the quasar feedback eventually removes most of the
gas (and dust), leaving the nucleus shining as an optical quasar
until the reservoir mass is exhausted on a timescale $\sim
10^7\,$yr (see Fig. \ref{bhlt}). Since then the host galaxy
evolves passively and the BH becomes dormant (apart from possible
reactivations). The process slows down with decreasing halo mass,
as the supernova feedback becomes increasingly important.}

\subsection{Photometric properties and metal abundances of spheroidal galaxies}

In small halos ($M_{\rm vir}\lesssim 10^{11}\,M_{\odot}$) the star
formation rate is kept low by stellar feedback, which heats up
most of the gas and  moves it to  lower density outskirts, where
the cooling time is very long.  Only when the virial mass exceeds
a few $10^{11}\,M_{\odot}$ the potential wells are deep enough to
allow a more effective star formation. Figure~\ref{MdmsuMsph}
shows the dependence on $M_{\rm vir}$ of the ratio $M_{\rm
vir}/M_{\rm sph}$, for three values of $z_{\rm vir}$. Bright
galaxies virializing at $z\lesssim 4$ are predicted to have
$M_{\rm vir}/M_{\rm sph}\simeq 40$. For comparison, McKay et al.
(2002) report, for bright galaxies, $M/L$ values in the SDSS
$g$-band of $171\pm 40$ based on the dynamics of satellites, and
of $270\pm 35$ based on weak lensing measurements; adopting for
the stellar component of early-type galaxies $M/L_V \simeq 4$--5.9
(Fukugita et al. 1998) this translates (for $L_V\simeq L_g$) into
$M_{\rm vir}/M_{\rm sph}\simeq 30$--70.

\begin{figure}
\epsscale{1.} \plotone{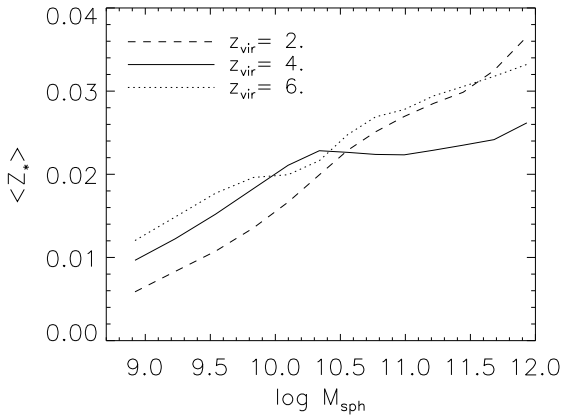} \caption{Mean stellar metallicity
as a function of the present-day stellar mass for the usual 3
values of $z_{\rm vir}$.} \label{metallicity}
\end{figure}

\begin{figure}
\plotone{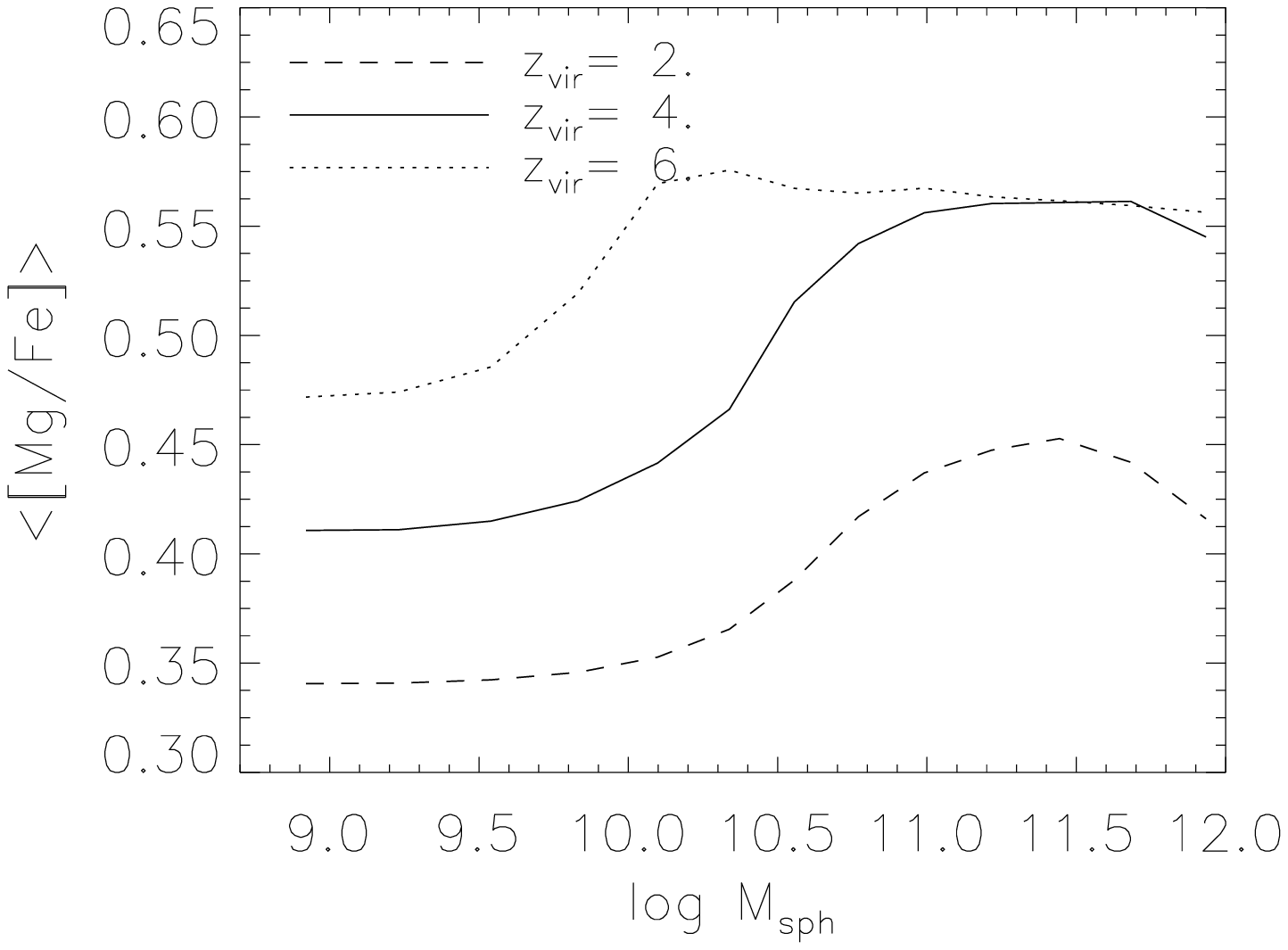} \caption{Mean $Mg/Fe$ abundance ratio as a
function of the present-day stellar mass for the usual 3 values of
$z_{\rm vir}$.} \label{MgFe}
\end{figure}

\new{Once the star formation law is specified, the chemical
evolution of the galaxy is followed by using classical equations
and stellar nucleosynthesis prescriptions (Granato et al. 2001).
The predicted mean metallicity, $\langle Z_\star\rangle$ in stars
is shown, as a function of the mass in stars, in
Fig.~\ref{metallicity}. The average metallicity increases from
about half solar to super-solar with increasing galaxy mass. In
particular, for our reference model, the metal content is
super-solar for $M_{\rm vir} \gtrsim 5 \times 10^{11}\,
M_{\odot}$, or $M_{\rm sph} \gtrsim 10^{10} \, M_{\odot}$.}

It should be stressed that this result refers to the average
metallicity of the galaxies, since our model is one-zone. In the
central regions, star-formation, and consequently metal
enrichment, is faster because of the higher densities entailing
shorter dynamical times. Adopting densities appropriate for the
innermost 10\% of the galaxy mass, the predicted metallicity of
the gas is between 3 to 5 times solar, with a general trend at
increasing with $M_{\rm vir}$. This result compares fairly well
with several recent {\bf direct} chemical measurements of
circum-nuclear gas in high-z QSO (e.g.\ Dietrich et al.\ 2003).

\new{Another evident trend is the increase with the galaxy mass of
the ratio between the abundance of the so-called $\alpha$-elements
and that of $Fe$. Fig.~\ref{MgFe} shows the mean $[Mg/Fe]$ ratio,
in stars, as a function of the galaxy mass for three different
values of the virialization redshift. The ratio $[Mg/Fe]$ is
always super-solar, and is further enhanced for $M_{\rm vir}
\gtrsim 10^{12}\,M_\odot$. However, in spite of the high
metallicity reached by our models, we find that the average
stellar  $[Fe/H ]$ is always subsolar ($\leq -0.25$). This trend
is dictated by the mass dependence of $T_{100}$: in high mass
objects the star-formation activity is essentially ended when SNIa
start to pollute the ISM with $Fe$.

Both the trend of the global metallicity and the ratio of the
$\alpha$- elements to $Fe$ abundance with the galaxy mass, are in
fair agreement with what is inferred from the spectrophotometric
observations of local early type galaxies. The narrowness and the
inclination of the color--magnitude relation of elliptical
galaxies hints at a rapid formation process and at a range in
metallicity from about half solar to twice solar (e.g. Bressan,
Chiosi \& Fagotto 1994). Furthermore, narrow band observations not
only provide evidence for a similar spread in metallicity
(Bernardi et al. 2003b), but also demand a significant enhancement
of the $\alpha$-elements in the more massive ellipticals (Worthey
et al. 1994). The observed run of $Mg$ and $Fe$ narrow band
indices with the luminosity of the galaxy or with the $H\beta$
index, cannot be reproduced by  stellar population models based
solely on solar partition of heavy elements (Worthey et al. 1994;
Trager et al. 2000a,b; Bernardi et al. 2003b).

It is worth noticing at this level that while the model
reproduces, for a relatively broad range of parameter choices, the
main trends of the metal abundances inferred  from observations of
local ellipticals, our predictions cannot yet be directly compared
with observations. For example, narrow band indices show a
complicated dependence on global metallicity, partition of heavy
elements and age of the stellar populations, that render the
adoption of common scaling relations (e.g. Matteucci, Ponzone \&
Gibson 1998) quite uncertain. We should thus generate mock
catalogues of local early type galaxies and investigate their
spectrophotometric properties by means of adequate spectral
synthesis tools. Work in this direction is in progress and will be
reported in a subsequent paper.}

\begin{figure}[t]
\epsscale{1.} \plotone{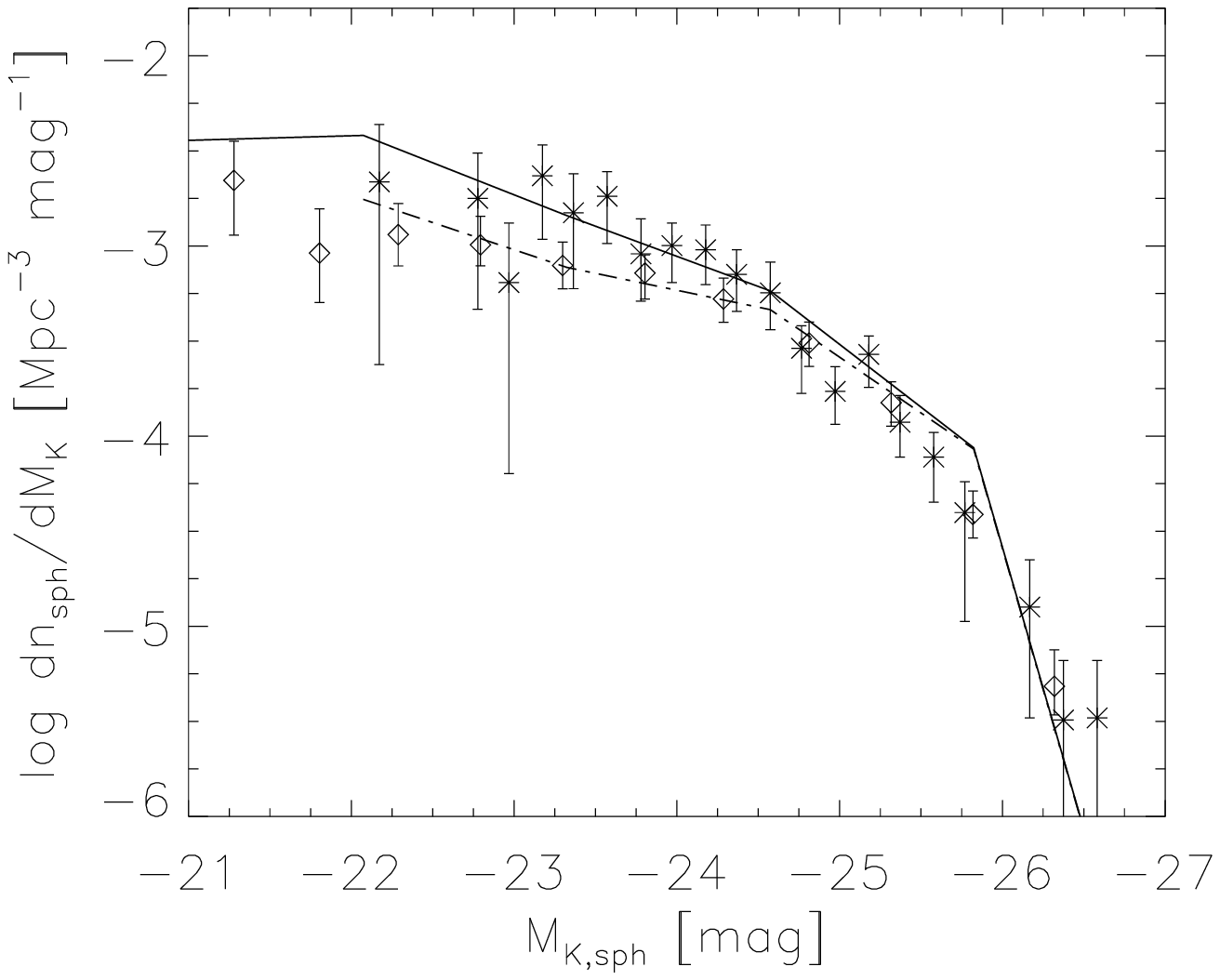} \caption{Predicted K-band local
luminosity function of massive spheroids (solid line) compared
with observational determinations by Kochanek et al. (2001;
diamonds) and Huang et al. (2003; stars). The dot-dashed line
shows the predicted local luminosity function of early-type
galaxies only, i.e. after subtracting from the global luminosity
function of spheroids the contributions of bulges of Sa and Sb
galaxies (see text). } \label{KLF}
\end{figure}

\subsection{Luminosity function of early-type galaxies}

In these galaxies the K-band luminosity is a quite good indicator
of the mass in stars. In Fig.~\ref{KLF} we compare the modelled
K-band local luminosity function with the recent determinations by
Kochanek et al. (2001) and Huang et al. (2003). \new{As already
remarked,} the solid line includes the contribution of all
spheroids with mass in the range $2.5 \times 10^{11}\lesssim
M_{\rm vir}\lesssim 1.6 \times 10^{13}$, formed at $z_{\rm vir}
\gtrsim 1.5$. According to our model, spheroidal galaxies born in
halos with $M_{\rm vir}\lesssim 2.5 \times 10^{11}\, M_{\odot}$
have masses in stars $M_{\rm sph}\lesssim 10^9\, M_{\odot}$ and
K-band magnitudes $\gtrsim -21.3$ (cfr. Fig.~\ref{MdmsuMsph}).

For a more direct comparison with the observational data, the
dot-dashed line in Fig.~\ref{KLF} shows the predicted local
luminosity function for early-type galaxies only. It was obtained
subtracting from the results shown by the solid line, the
contributions of bulges of Sa and Sb galaxies estimated using the
local B-band luminosity functions and the bulge-to-total
luminosity ratios for galaxies of these morphological types
derived by Salucci et al. (1999) and adopting a color $B-K=4.1$.
It should also be noted that the mass in stars for small halos is
strongly dependent on the stellar feedback, namely on the IMF and
on the SN efficiency $\epsilon_{\rm SN}$: an increase of the
efficiency from 0.1 to 0.3 results in a decrease of mass in stars
by a factor $\simeq 3$. This dependence weakens with increasing
$M_{\rm vir}$.

\begin{figure}[t]
\epsscale{1.} \plotone{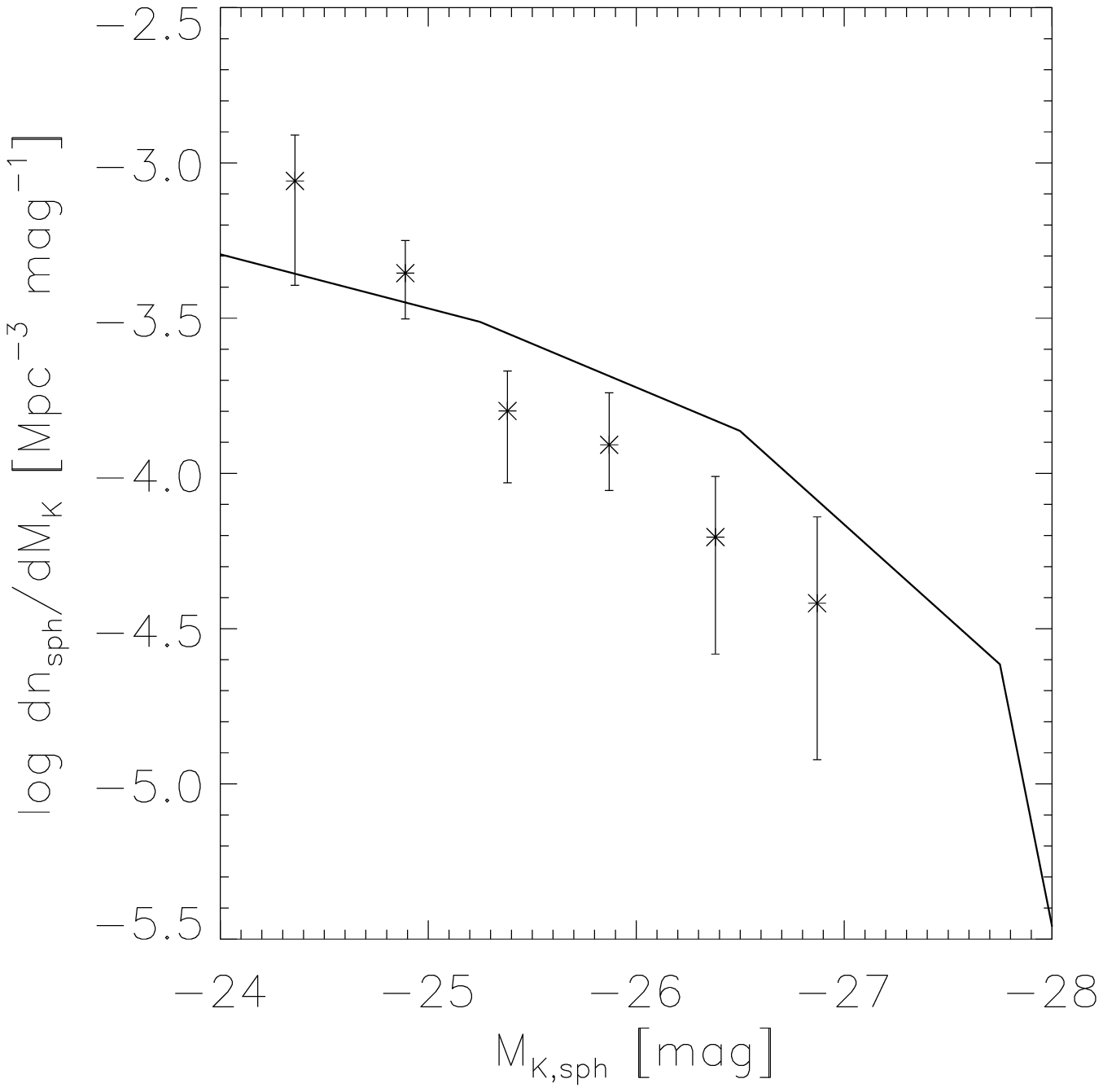} \caption{Predicted K-band
luminosity function of massive spheroids at $z=1.5$ compared with
observational determination by Pozzetti et al. (2003). }
\label{Pozzetti}
\end{figure}

Figure~\ref{Pozzetti} compares the predicted rest-frame K-band
luminosity function at $z = 1.5$ with the observational
determination by Pozzetti et al. (2003), which, at the highest
luminosities, is substantially above predictions of the models by
Kauffmann et al. (1999), Cole et al. (2000), and Menci et al.
(2002).



\begin{figure}[t]
\epsscale{1.} \plotone{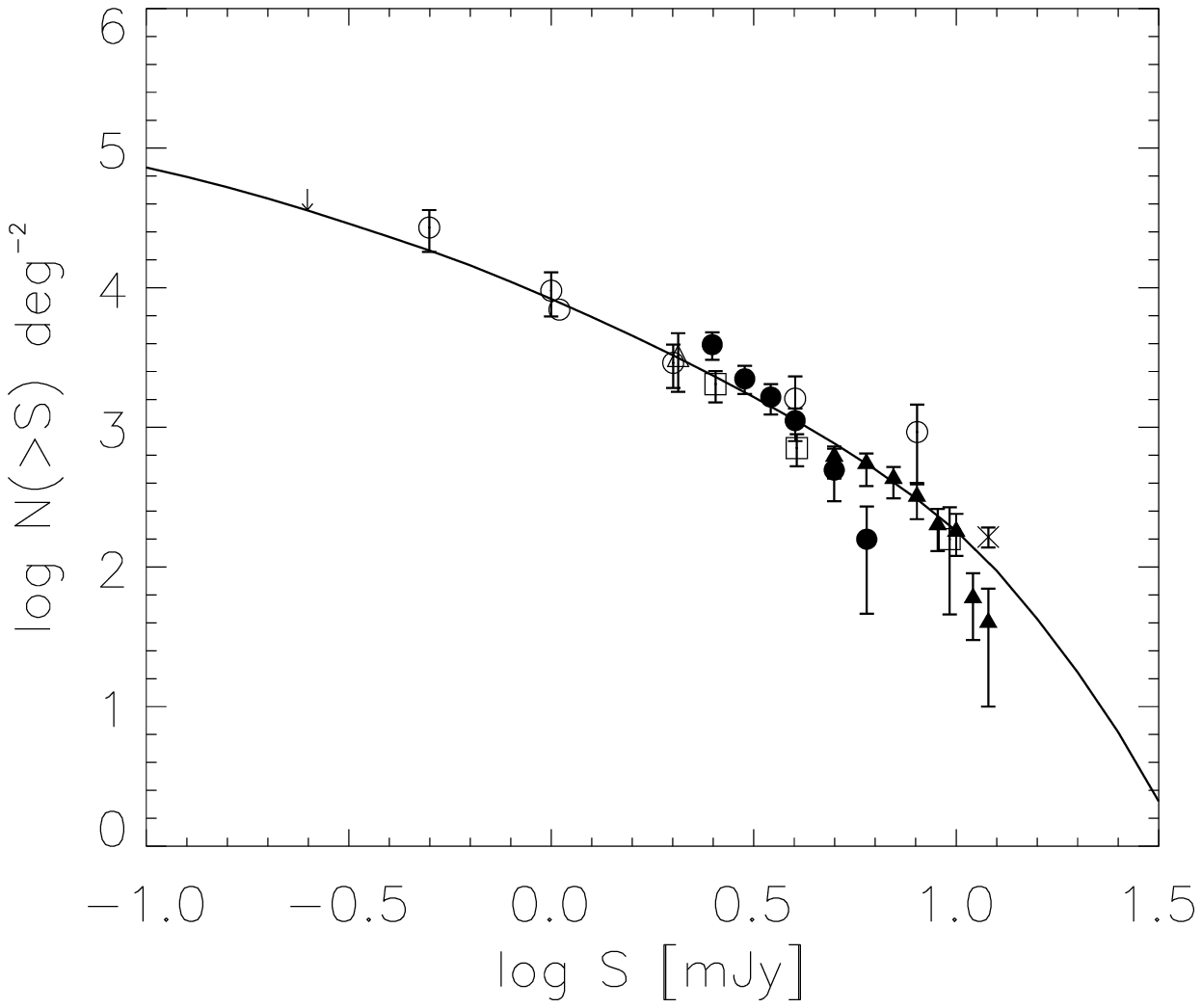} \caption{Predicted $850\,\mu$m
extragalactic counts compared with SCUBA counts by Blain et al.
(1999, open circles), Hughes et al. (1998, star and triangles),
Barger et al. (1999, open squares), Eales et al. (2000, filled
circles), Chapman et al. (2002, filled triangle), and Borys et al.
(2002, filled squares). } \label{SCUBAcounts}
\end{figure}

\begin{figure}[t]
\epsscale{1.} \plotone{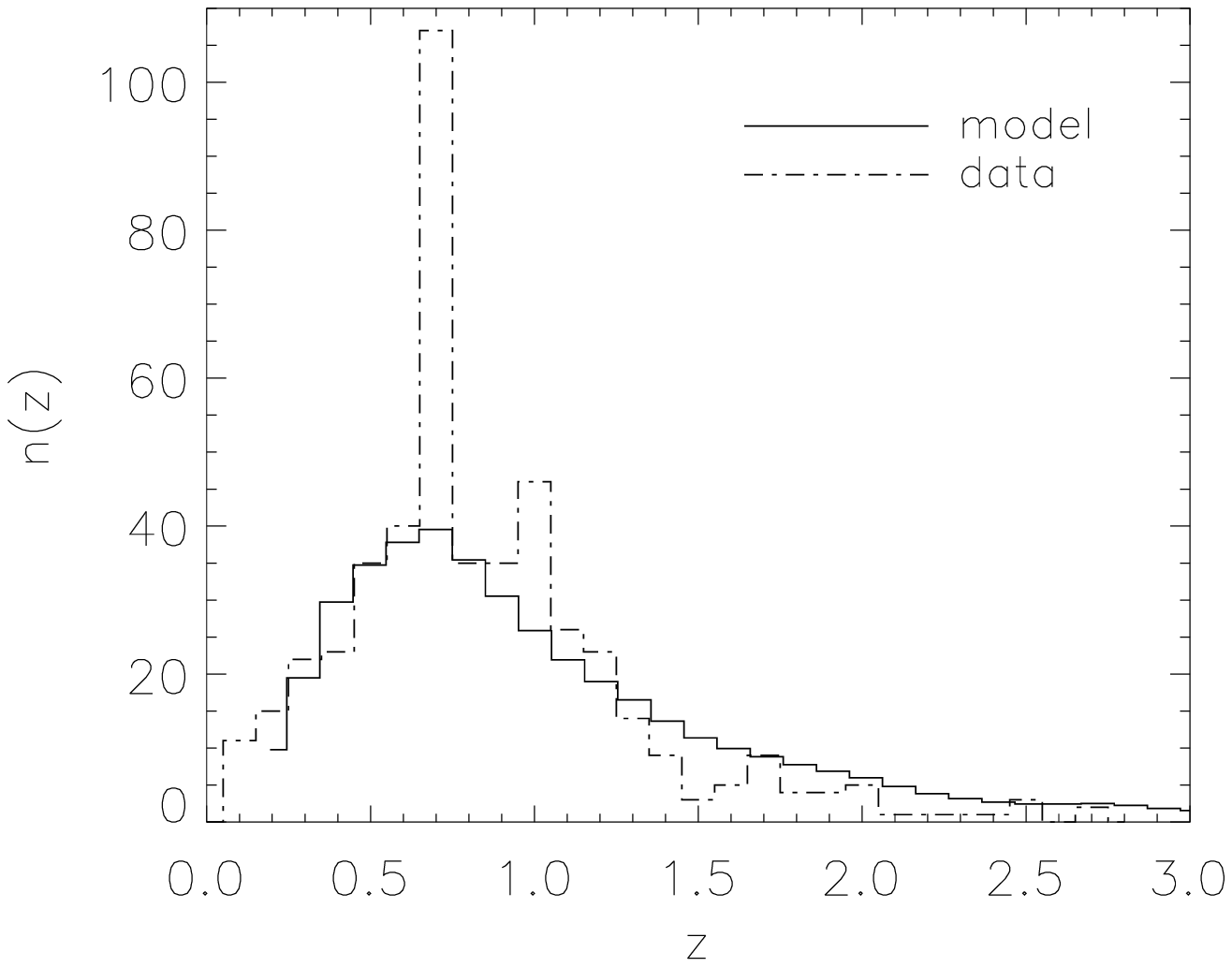} \caption{Predicted redshift
distribution of galaxies brighter than $K=20$ compared with the
results of the K20 survey (Cimatti et al. 2002b). }
\label{zdistrK20}
\end{figure}

\subsection{SCUBA galaxies and EROs}

In large galactic halos the star formation rate turns out to be
very high at high redshift, yielding a quick increase of the
metallicity and of  the dust mass. The latter is computed by
GRASIL as proportional to the product of the gas mass by its
metallicity, with a coefficient determined by the condition of a
gas-to-dust ratio of 110 for solar metallicity (Silva et al.\
1998). Thus, most of the star formation occurs in a dusty
environment, so that these galaxies are powerful far-IR/sub-mm
sources, highly obscured in the visual and near IR bands. In
Fig.~\ref{SCUBAcounts} we have plotted the model predictions at
$850\,\mu$m against the SCUBA counts. The predicted redshift
distributions are similar to those shown by Granato et al. (2001).
In particular, for a flux density limit of 5 mJy, the model gives
a median redshift of 2.2 and an inter-quartile range of 1.6--3.3,
to be compared with $z_{\rm median} =2.4$ and the inter-quartile
range of 1.9--2.8 found for the sample of Chapman et al. (2003).

After the interstellar medium has been swept out, galaxies evolve
passively. The combination of redshift and aging soon makes them
extremely red. We computed the expected contribution to the
extragalactic K-band counts of spheroidal galaxies in this phase.
A comparison (Fig.~\ref{zdistrK20}) of the predicted with the
observed redshift distribution of galaxies with $K\leq 20$
(Cimatti et al. 2002b) shows that they fully saturate the high
redshift tail of the distribution.

Cimatti et al. (2002a) selected a complete sample of Extremely Red
Objects (EROs) ($(R-K_s)\geq 5$) with $K< 19.2$. More than  60$\%$
of the objects have redshift, mostly spectroscopic. On the basis
of the spectra, the sample has been subdivided into dusty and
non-dusty EROs. Their data suggest that there is a significant
number of {\it old} dust-free ellipticals in place at z$\geq 1$.
This result is confirmed by the subsequent analysis by Pozzetti et
al. (2003), who found that the bright end of the K-band luminosity
function at $z\geq 1$ is dominated by red/early type galaxies. Our
model is consistent with these results (cf. Fig.~\ref{Pozzetti}).

Our expectation of the existence of a significant population of
luminous red galaxies at substantial redshifts is also borne out
by the results of the very deep near-infrared photometry of the
Hubble Deep Field-South with ISAAC on the VLT (Franx et al. 2003)
and by follow-up Keck spectroscopy (van Dokkum et al. 2003). Model
predictions closely match the surface density ($3\pm
0.8\,\hbox{arcmin}^{-2}$) and the median redshift (2.6) of
galaxies with $K_s<22.5$ and $(J_s-K_s)>2.3$ (Franx et al. 2003).

\section{Discussion}

To highlight the dependence of galaxy properties on their halo
masses, we make reference to three characteristic values, namely
$M_{\rm vir}\simeq 10^{11.4}$, $10^{12.4}$ and $10^{13.4}$,
referred to as low, intermediate and high masses, respectively.

The clumping factor $C$ affects only the evolution of high and
intermediate masses, where a decrease of $C$ from $20$ to 1
strongly inhibits the capability of gas to quickly cool down, form
stars and feed the BH. For large galactic masses, the
star-formation activity and the growth of the BH, rather than
being confined to the first $\sim 1\,$Gyr after virialization,
continues for a time comparable to the Hubble time. However, the
final ratio $M_{\rm BH}/M_{\rm sph}$ is almost the same, both
masses ultimately depending on the SFR. Conversely, at lower
masses the collapse becomes increasingly limited by the dynamical
time even when $C=1$.

The supernova efficiency $\epsilon_{\rm SN}$ has some effect on
the evolution at all  masses: $M_{\rm sph}$, $<Z_{*}>$ and $M_{\rm
BH}/M_{\rm sph}$ decrease with increasing $\epsilon_{\rm SN}$,
while $T_{100}$ increases. Differences are modest at high masses,
but become dramatic at low masses. At $M_{\rm vir} \lesssim 2
\times 10^{11}\,M\odot$ and $\epsilon_{\rm SN}\gtrsim 0.2$, the
evolution of the star-formation rate can show damped oscillations
at early epochs ($T\lesssim 1\,$Gyr).

By converse, the QSO efficiency, $\epsilon_{\rm QSO}$, influences
only intermediate and large masses, where an increase of this
parameter determines a shorter active star forming phase, and
lower values of $\langle Z_{*}\rangle$, $M_{\rm sph}$ and $M_{\rm
QSO}$, while the ratio $M_{\rm BH}/M_{\rm sph}$ is little
affected. An increase of the radiative efficiency $\epsilon$ has a
very similar effect.

When the accretion is Eddington-limited (or, at most, mildly
super-Eddington), $T_{100}$, $\langle Z_{*}\rangle$, and $M_{\rm
sph}$ decrease with increasing seed BH mass: less e-folding times
are necessary to produce substantial effects on the environment.
Typically, $M_{\rm sph}$ decreases by 30\% when the seed mass is
increased from $10^3$ to $10^4 M_\odot$, but the ratio $M_{\rm
BH}/M_{\rm sph}$ decreases by less than 10\%. As already remarked,
the development of low masses is instead weakly affected by the BH
feedback. By converse, if substantially super-Eddington accretion
is allowed, the model is quite insensitive to the precise choice
of the seed BH mass.

\begin{figure}[t]
\epsscale{1.} \plotone{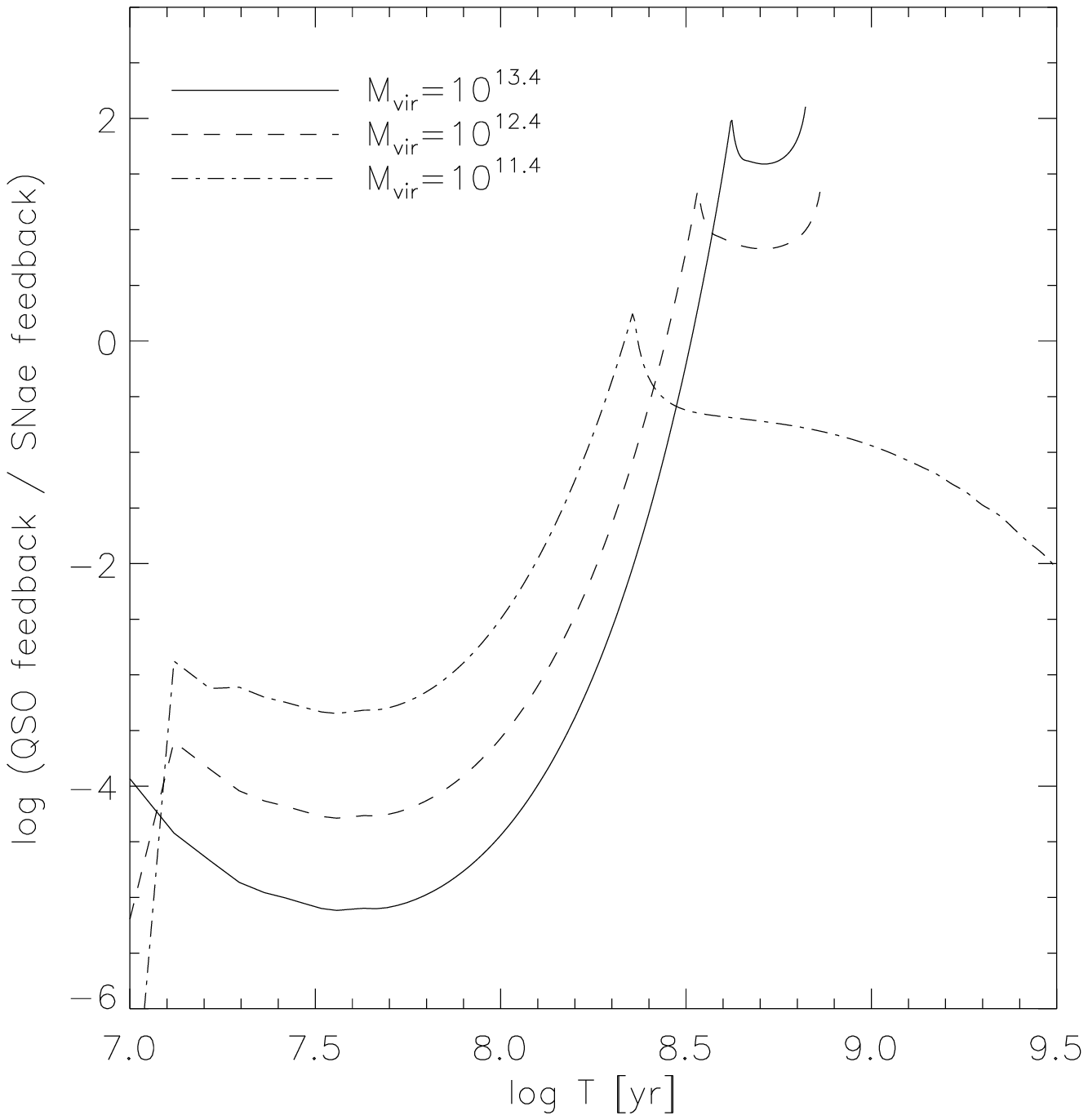} \caption{Ratio between the mass
ejected from the galaxy per unit time by the feedback of the QSO
and by that of SNae.
 } \label{fig:ratiowinds}
\end{figure}

\new{ The ratio between QSO and SN feedback is an increasing
function of the virial mass of the galaxy (see
Fig.~\ref{fig:ratiowinds}). For low mass galaxies the integrated
effect of the QSO is almost negligible compared to that of SNae,
but takes over (typically by a factor of a few) for intermediate
masses ($M_{\rm vir}\sim 10^{12.4}\, M_\odot$), and dominates (by
a factor $\gtrsim 10$) at high masses ($M_{\rm vir}\sim
10^{13.4}\, M_\odot$). Note that the QSO effect usually increases
exponentially with time, while that of SNae increases more slowly.
Thus the instantaneous QSO effect becomes dominant, if ever, only
a few e-folding times before the maximum of QSO activity. }

\new{As mentioned at the beginning of Sect.~3, our definition of a
galactic halo associated with a spheroidal galaxy is rather crude.
However, the successful comparison with the observed population
properties of spheroidal galaxies (Figs.~\ref{BHmassfunction},
\ref{KLF}, \ref{Pozzetti}, \ref{SCUBAcounts}, \ref{zdistrK20}, and
\ref{cosmic}) gives us some confidence about the meaningfulness of
the criterion we have adopted.

Although we have not addressed the details of the formation of
disk (and irregular) galaxies, we envisage them as associated
primarily to halos virializing at $z_{\rm vir} \lesssim 1.5$,
which have incorporated, through merging processes, a large
fraction of halos less massive than $2.5 \times 10^{11}\,M_\odot$
virializing at earlier times. These low mass halos virialized at
early times may become the bulges of late type galaxies.}

\section{Summary and conclusions}

We have presented a detailed, physically grounded, model for the
early co-evolution of spheroidal galaxies and of active nuclei at
their centers. The model is based on very simple recipes, that can
be easily implemented.  In summary, we start from the diffuse gas
within the dark matter halo falling down into the star forming
regions at a rate ruled by the dynamic and the cooling times. Part
of this gas condenses into stars, at a rate again controlled by
the local dynamic and cooling times. But the gas also feels the
feedback from supernovae and from active nuclei, heating it and
possibly expelling it from the potential well. Also the radiation
drag on the cold gas decreases its angular momentum, causing an
inflow into a reservoir around the central black hole. Viscous
drag then causes the gas to flow from the reservoir into the black
hole, increasing its mass and powering the nuclear activity.

In the shallower potential wells (corresponding to lower halo
masses and, for given mass, to lower virialization redshifts), the
supernova heating is increasingly effective in slowing down the
star formation and in driving gas outflows, resulting in an
increase of star/dark-matter ratio with increasing halo mass. As a
consequence, the star formation is faster within the most massive
halos, and the more so if they virialize at substantial redshifts.
Thus, in keeping with the proposition by Granato et al. (2001),
physical processes acting on baryons effectively reverse the order
of formation of galaxies compared to that of dark-matter halos.

A higher star-formation rate also implies a higher radiation drag,
resulting in a faster loss of angular momentum of the gas (Umemura
2001; Kawakatu \& Umemura 2002; Kawakatu et al. 2003) and,
consequently, in a faster inflow towards the central black-hole.
In turn, the kinetic energy carried by outflows driven by active
nuclei through line acceleration is proportional to $M_{\rm
BH}^{3/2}$ (Murray et al. 1995), and this mechanism can inject in
the interstellar medium a sufficient amount of energy to unbind
it. The time required to sweep out the interstellar medium, thus
halting both the star formation and the black-hole growth, is
again shorter for larger halos. For the most massive galaxies
($M_{\rm vir}\gtrsim 10^{12}\,M_\odot$) virializing at $3 \lesssim
z_{\rm vir} \lesssim 6$, this time is $< 1\,$Gyr, so that the bulk
of the star-formation may be completed before type Ia supernovae
can significantly increase the $Fe$ abundance of the interstellar
medium; this process can then account for the $\alpha$-enhancement
seen in the largest galaxies.

The interplay between star formation and nuclear activity
determines the relationship between the black-hole mass and the
mass, or velocity dispersion, of the host galaxy, as well as the
black-hole mass function. As illustrated by Figs.~\ref{MBHsigma}
and \ref{BHmassfunction}, the model predictions are in excellent
agreement with the observational data. A specific prediction of
the model is a substantial steepening of the $M_{\rm
BH}$--$\sigma$ relation for $\sigma \lesssim
150\,\hbox{km}\,\hbox{s}^{-1}$: the mass of the BH associated to
less massive halos is lower than expected from an extrapolation
from higher masses, because of the combined effect of SN heating,
which is increasingly effective with decreasing galaxy mass in
hindering the gas inflow towards the central BH, and of the
decreased radiation drag [see Eq.~(\ref{dotMres}), with $\tau \ll
1$].

Coupling the model with GRASIL (Silva et al. 1998), the code
computing in a self-consistent way the chemical and
spectrophotometric evolution of galaxies over a very wide
wavelength interval, we have obtained predictions for the sub-mm
counts and the corresponding redshift distributions as well as for
the redshift distributions of sources detected by deep K-band
surveys, which proved to be extremely challenging for all the
current semi-analytic models. The results, shown by
Figs.~\ref{SCUBAcounts} and \ref{zdistrK20}, are again very
encouraging.

\begin{figure}[t]
\epsscale{1.} \plotone{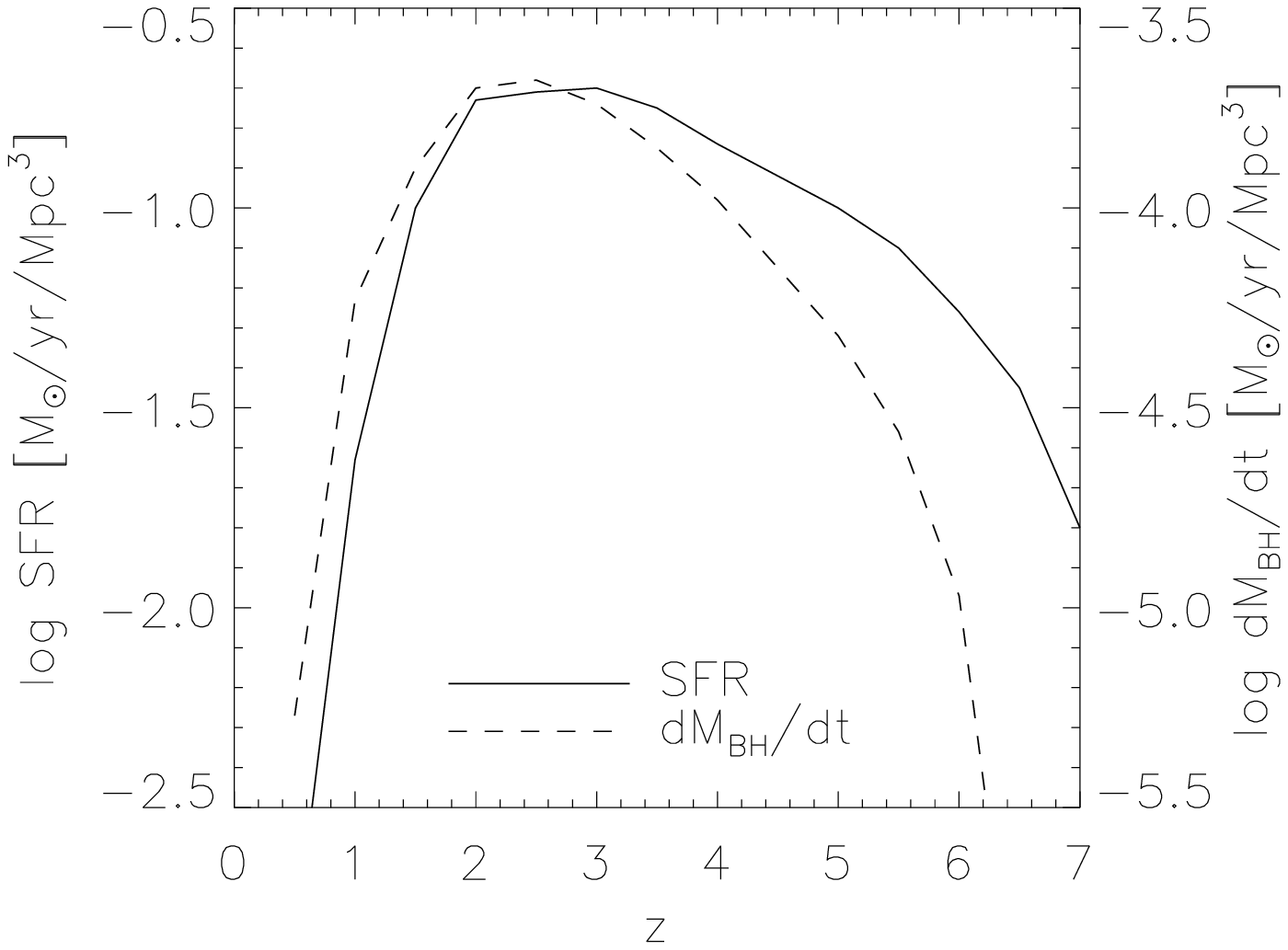} \caption{Predicted star formation
in spheroids and BH mass accretion rates per unit volume as a
function of redshift.} \label{cosmic}
\end{figure}

\new{A discussion of the evolutionary properties of AGNs predicted
by the present model is deferred to a future paper, where we will
address, among other things, the complex issue of how the
bolometric luminosity produced by accretion processes shows up in
different electromagnetic bands. We note however that the analysis
by Granato et al. (2001), who approached the problem the other way
round, i.e. inferred the formation history of spheroidal galaxies
and of galactic bulges from the observed epoch-dependent
luminosity function of quasars, got results in detailed agreement
with those presented here. Thus an at least qualitative agreement
of the present model with the data on AGNs seems to be ensured.
This is confirmed by the successful comparison of our model
predictions for the relationship among the black-hole mass and the
velocity dispersion of the host galaxy (Fig.~\ref{MBHsigma}), and
for the local black-hole mass function
(Fig.~\ref{BHmassfunction}). Also, the predicted history of global
accretion rate onto the central black-holes (Fig.~\ref{cosmic}),
which, in our scheme, is directly proportional to the history of
the bolometric luminosity density produced by AGNs, with its peak
in the redshift range 2 to 3, is nicely consistent with the
results of optical surveys (see, e.g., Fig. 8 of Fan et al.
2003).}

\noindent{\bf Acknowledgments} Work supported in part by MIUR and
ASI. We warmly thank the referee for her/his constructive
comments.


\begin{thebibliography}{}
\bibitem[{Arav} 1994]{1994ApJ...432...62A} Arav, N.,
Li, Z., \& Begelman, M.C. 1994, ApJ,  432, 62
\bibitem[]{} Aretxaga, I., Hughes, D.H., Chapin, E.L., Gaztanaga, E., Dunl
J.S.,\& Ivison, R. 2003, MNRAS, in press
\bibitem[{Barger} 1999]{1999ApJ...518L...5B} Barger, A.J.,
Cowie, L.L., \& Sanders, D.B. 1999, ApJ,  518, L5
\bibitem[{Begelman} 2001]{2001ApJ...551..897B}
Begelman, M.C. 2001, ApJ,  551, 897
\bibitem[{Begelman} 2002]{2002ApJ...568L..97B}
Begelman, M.C. 2002, ApJ,  568, L97
\bibitem[]{} Begelman, M.C. 2003, in Carnegie
Observatories Astrophysics Series, Vol. 1: Coevolution of Black
Holes and Galaxies, ed. L.C. Ho (Cambridge: Cambridge Univ.
Press), in press, astro-ph/0303040
\bibitem[{Bennett} 2003]{2003ApJ...583....1B} Bennett,
C.L., et al. 2003, ApJ,  583, 1
\bibitem[Bernardi et al.(2003)]{2003AJ....125.1817B} Bernardi, M.,~et al.\
2003a, \aj, 125, 1817
\bibitem[Bernardi et al.(2003b)]{2003AJ....125.1882B} Bernardi, M.,~et al.\
2003b, \aj, 125, 1882
\bibitem[{Bicknell} 1997]{1997ApJ...485..112B}
Bicknell, G.V., Dopita, M.A., \& O'Dea, C.P.O. 1997, ApJ,  485,
112
\bibitem[{Blain} 1999]{1999ApJ...512L..87B} Blain, A.W.,
Kneib, J.-P., Ivison, R.J., \& Smail, I. 1999, ApJ,  512, L87
\bibitem[{Blain} 2002]{2002PhR...369..111B} Blain, A.W.,
Smail, I., Ivison, R.J., Kneib, J.-P., \& Frayer, D.T., 2002, PhR,
369, 111
\bibitem[{Borys} 2002]{2002MNRAS.330L..63B} Borys, C.,
Chapman, S.C., Halpern, M., \& Scott, D. 2002, MNRAS,  330, L63
\bibitem[]{}
Bower, R.G., Benson, A.J., Lacey, C.G., Baugh, C.M., Cole, S., \&
Frenk, C.S. 2001, MNRAS, 325, 497
\bibitem[{Brandt} 2000]{2000NewAR..44..461B}
Brandt, W.N., \& Gallagher, S.C. 2000, NewAR,  44, 461
\bibitem[{Brandt} 2001]{2001ncxa.conf..128B} Brandt, W.N.,
Gallagher, S.C., \& Kaspi, S. 2001, in New Century of X-ray
Astronomy, ed. H. Inoue \& H. Kunieda (ASP Conf. Ser. 251; S.
Francisco: ASP), 128
\bibitem[\protect\citeauthoryear{Bressan, Chiosi, \&
Fagotto}{1994}]{1994ApJS...94...63B} Bressan, A., Chiosi, C., \&
Fagotto, F. 1994, ApJS, 94, 63
\bibitem[{Br{\" u}ggen} 2002]{2002MNRAS.331..545B} Br{\"
u}ggen, M., Kaiser, C.R., Churazov, E., En{\ss}lin, T.A. 2002,
MNRAS, 331, 545
\bibitem[]{} Bullock, J.S., Dekel, A., Kolatt, T.S., Kravtsov, A.V.,
Klypin, A.A., Porciani, C., \& Primack, J.R. 2001a, ApJ,  555, 240
\bibitem[{Bullock} 2001]{2001MNRAS.321..559B}
Bullock, J.S., Kolatt, T.S., Sigad, Y., Somerville, R.S.,
Kravtsov, A.V., Klypin, A.A., Primack, J.R., \& Dekel, A. 2001b,
MNRAS,  321, 559
\bibitem[{Bullock} 2002]{2002MNRAS.329..246B}
Bullock, J.S., Wechsler, R.H., \& Somerville, R.S. 2002, MNRAS,
329, 246
\bibitem[{Burkert} 2001]{2001ApJ...554L.151B}
Burkert, A., \& Silk, J. 2001, ApJ,  554, L151
\bibitem[{Cavaliere} 2002]{2002ApJ...581L...1C}
Cattaneo, A., \& Bernardi, M., 2003, MNRAS, in press
\bibitem[{Cavaliere} 2002]{2002ApJ...581L...1C}
Cavaliere, A., Lapi, A., \& Menci, N. 2002, ApJ,  581, L1
\bibitem[{Cavaliere} 1998]{1998yugf.conf...26C}
Cavaliere, A., \& Vittorini V. 1998, in The Young Universe: Galaxy
Formation and Evolution at Intermediate and High Redshift, ed. S.
D'Odorico, A. Fontana, \& E. Giallongo (ASP Conf. Ser. 146; S.
Francisco: ASP), 26
\bibitem[]{}
Celotti, A., Padovani, P., \& Ghisellini, G. 1997, MNRAS, 286, 415
\bibitem[{Chapman} 2003]{2003Natur.422..695C}
Chapman, S.C., Blain, A.W., Ivison, R.J., \& Smail, I.R. 2003,
Nat,  422, 695
\bibitem[{Chapman} 2002]{2002MNRAS.330...92C}
Chapman, S.C., Scott, D., Borys, C., \& Fahlman, G.G. 2002, MNRAS,
330, 92
\bibitem[]{}
Chartas, G., Brandt, W.N., \& Gallagher, S.C. 2003, ApJ in press,
astro-ph/0306125
\bibitem[]{}
Chartas, G., Brandt, W.N., Gallagher, S.C., \& Garmire, G.P. 2002,
ApJ, 579, 169
\bibitem[{Cimatti} 2002]{2002A&A...381L..68C} Cimatti, A.,
et al. 2002a, A\&A,  381, L68
\bibitem[{Cimatti} 2002]{2002A&A...391L...1C} Cimatti, A.,
et al. 2002b, A\&A,  391, L1
\bibitem[{Cole} 2000]{2000MNRAS.319..168C} Cole, S.,
Lacey, C.G., Baugh, C.M., \& Frenk, C.S. 2000, MNRAS,  319, 168
\bibitem[{Dietrich} 2002]{2002ApJ...564..581D}
Dietrich, M., Hamman, F., Schields, J.C., Constantin, A., Heidt,
J., J\"ager, K., Vestergaard, M., \& Wagner, S.J. 2003, ApJ, 589,
722
\bibitem[{Devriendt} 2000]{2000A&A...363..851D}
Devriendt, J.E.G., \& Guiderdoni, B., 2000 A\&A,  363, 851
\bibitem[]{}
Di Matteo, T., Croft, R.A.C., Springel, V., \& Hernquist, L.,
2003, ApJ, 593, 56
\bibitem[{Dunlop} 2001]{2001NewAR..45..609D} Dunlop, J.S.
2001, NewAR,  45, 609
\bibitem[{Dunlop} 2003]{2003MNRAS.340.1095D} Dunlop, J.S.,
McLure, R.J., Kukula, M.J., Baum, S.A., O'Dea, C.P., Hughes, D.H.
2003, MNRAS,  340, 1095
\bibitem[{Duschl} 2000]{2000A&A...357.1123D} Duschl, W.J.
Strittmatter, P.A., \& Biermann, P.L. 2000, A\&A,  357, 1123
\bibitem[{Eales} 2000]{2000AJ....120.2244E} Eales, S.,
Lilly, S., Webb, T., Dunne, L., Gear, W., Clements, D., \& Yun, M.
2000, AJ,  120, 2244
\bibitem[]{}
Egami, E., Neugebauer, G., Soifer, B.T., Matthews, K., Ressler,
M., Becklin, E.E., Murphy, T.W. Jr., \& Dale, D.A. 2000, ApJ, 535,
561
\bibitem[{Eisenhauer} 2001]{2001slbt.work...89E}
Eisenhauer, F. 2001, in Science with the Large Binocular
Telescope, ed. T. Herbst, (Neumann Druck), 89
\bibitem[\protect\citeauthoryear{Elvis, Risaliti, \&
Zamorani}{2002}]{2002ApJ...565L..75E} Elvis, M., Risaliti, G., \&
Zamorani, G. 2002, ApJ, 565, L75
\bibitem[{En{\ss}lin} 2000]{2000A&A...360..417E}
En{\ss}lin, T.A., \& Kaiser, C.R. 2000, A\&A,  360, 417
\bibitem[{Fabian} 1999]{1999MNRAS.308L..39F} Fabian, A.C.
1999, MNRAS,  308, L39
\bibitem[\protect\citeauthoryear{Fan et al.}{2000}]{2000AJ....120.1167F}
Fan, X., et al. 2000, AJ, 120, 1167
\bibitem[\protect\citeauthoryear{Fan et al.}{2001}]{2001AJ....122.2833F}
Fan, X., et al. 2001, AJ, 122, 2833
\bibitem[\protect\citeauthoryear{Fan et al.}{2003}]{2003AJ....125.1649F}
Fan, X., et al. 2003, AJ, 125, 1649
\bibitem[{Ferrarese} 2002]{2002ApJ...578...90F}
Ferrarese, L. 2002, ApJ,  578, 90
\bibitem[]{}
Ferrarese, L., \& Merritt, D. 2000, ApJ, 539, L9
\bibitem[{Forbes} 1999]{1999MNRAS.309..623F}
Forbes, D.A., \& Ponman, T.J. 1999, MNRAS,  309, 623
\bibitem[{Franx} 2003]{2003ApJ...587L..79F} Franx, M.,
et al. 2003, ApJ,  587, L79
\bibitem[{Freudling} 2003]{2003ApJ...587L..67F}
Freudling, W., Corbin, M.R., \& Korista, K.T. 2003, ApJ,  587, L67
\bibitem[]{} Fria\c{c}a, A.C.S., \& Terlevich, R.J. 1998, MNRAS, 298,
399
\bibitem[{Furlanetto} 2001]{2001ApJ...556..619F}
Furlanetto, S.R., \& Loeb, A. 2001, ApJ,  556, 619
\bibitem[]{}
Gebhardt, K., et al. 2000, ApJ, 539, L13
\bibitem[{Granato} 2000]{2000ApJ...542..710G}
Granato, G.L., Lacey, C.G., Silva, L., Bressan, A., Baugh, C.M.,
Cole, S., \& Frenk, C.S. 2000, ApJ,  542, 710
\bibitem[]{}
Granato, G.L., Silva, L., Monaco, P., Panuzzo, P., Salucci, P., De
Zotti, G., \& Danese, L. 2001, MNRAS, 324, 757
\bibitem[{Heckman} 2000]{2000ApJS..129..493H}
Heckman, T.M., Lehnert, M.D., Strickland, D.K., \& Armus, L. 2000,
ApJS, 129, 493
\bibitem[{Haehnelt} 2000]{2000MNRAS.318L..35H}
Haehnelt, M.G., \& Kauffmann, G. 2000, MNRAS,  318, L35
\bibitem[{Haehnelt} 1998]{1998MNRAS.300..817H}
Haehnelt, M.G., Natarajan P. \& Rees, M.J. 1998, MNRAS,  300, 817
\bibitem[{Haehnelt} 1993]{1993MNRAS.263..168H}
Haehnelt, M.G., \& Rees, M.J. 1993, MNRAS,  263, 168
\bibitem[]{} Haiman, Z., Ciotti, L., \& Ostriker, J.P., 2003, ApJ, submitt
astro-ph/0304129
\bibitem[]{}
Hamann, F., \& Ferland, G. 1999, ARA\&A, 37, 487
\bibitem[{Huang} 2003]{2003ApJ...584..203H} Huang, J.-S.,
Glazebrook, K., Cowie, L.L., \& Tinney, C. 2003, ApJ,  584, 203
\bibitem[]{}
Hughes, D.H., et al. 1998, \nat, 394, 241
\bibitem[{Hutchings} 2002]{2002AJ....123.2936H}
Hutchings, J.B., Frenette, D., Hanisch, R., Mo, J., Dumont, P.J.,
Redding, D.C., \& Neff, S.G. 2002, AJ,  123, 2936
\bibitem[{Inoue} 2001]{2001ApJ...562..618I}
Inoue, S., \& Sasaki, S. 2001, ApJ,  562, 618
2002, ApJ, 565, 63
\bibitem[]{} Ivison, R.J., et al. 2002, MNRAS,  337, 1
\bibitem[{Kashikawa} 2003]{2003AJ....125...53K}
Kashikawa, N., et al. 2003, AJ,  125, 53
\bibitem[{Kauffmann} 1999]{1999MNRAS.303..188K}
Kauffmann, G., Colberg, J.M., Diaferio, A., \& White, S.D.M. 1999,
MNRAS, 303, 188
\bibitem[{Kauffmann} 2000]{2000MNRAS.311..576K}
Kauffmann, G., \& Haehnelt, M. 2000, MNRAS,  311, 576
\bibitem[{Kawakatu} 2002]{2002MNRAS.329..572K}
Kawakatu, N., \& Umemura, M. 2002, MNRAS,  329, 572
\bibitem[{Kawakatu} 2003]{2003ApJ...583...85K}
Kawakatu, N., Umemura, M., \& Mori, M. 2003, ApJ,  583, 85
\bibitem[{Kochanek} 2001]{2001ApJ...560..566K}
Kochanek, C.S., et al. 2001, ApJ,  560, 566
\bibitem[{Kormendy} 2001]{2001tsra.conf..363K} Kormendy,
J., \& Gebhardt, K., 2001, in 20th Texas Symposium on relativistic
astrophysics, ed. J.C. Wheeler, \& H. Martel (AIP Conf. Proc. 586;
Melville: AIP), 363
\bibitem[]{}
Kormendy, J., \& Richstone, D. 1995, ARA\&A, 33, 581
\bibitem[]{}
Kravtsov, A.V., \& Yepes, G. 2000, MNRAS, 318, 227
\bibitem[{Kukula} 2001]{2001MNRAS.326.1533K} Kukula M.J.,
Dunlop, J.S., McLure, R.J., Miller, L., Percival, W.J., Baum,
S.A., \& O'Dea, C.P. 2001, MNRAS,  326, 1533
\bibitem[]{}
Laor, A., \& Brandt, W.N. 2002, ApJ, 569, 641
\bibitem[]{} Lapi, A., Cavaliere, A., \& De Zotti, G. 2003, ApJ,
submitted
\bibitem[{Larson} 1998]{1998MNRAS.301..569L} Larson, R.B.
1998, MNRAS,  301, 569
\bibitem[{Li} 2002]{2002ApJ...566..652L} Li, L.,
\& Ostriker, J.P. 2002, ApJ,  566, 652
\bibitem[\protect\citeauthoryear{Mac Low \&
Ferrara}{1999}]{1999ApJ...513..142M} Mac Low, M., \& Ferrara, A.,
1999, ApJ, 513, 142
\bibitem[]{}
Magliocchetti, M., Moscardini, L., Panuzzo, P., Granato, G.L., De
Zotti, G.,  \& Danese, L. 2001, MNRAS, 325, 1553
\bibitem[]{}
Magorrian, J., et al. 1998, AJ, 115, 2285
\bibitem[{Maller} 2002]{2002MNRAS.335..487M} Maller,
A.H., \& Dekel, A. 2002, MNRAS,  335, 487
\bibitem[]{}
Marconi, A., \& Hunt, L. 2003, ApJL, 589, L21
\bibitem[\protect\citeauthoryear{Matteucci, Ponzone, \&
Gibson}{1998}]{1998A&A...335..855M} Matteucci, F., Ponzone, R., \&
Gibson, B.K. 1998, A\&A, 335, 855
\bibitem[{McLure} 2002]{2002MNRAS.331..795M}
McLure, R.J., \& Dunlop, J.S. 2002, MNRAS,  331, 795
\bibitem[{Menci} 2002]{2002ApJ...575...18M} Menci, N.,
Cavaliere, A., Fontana, A., Giallongo, E., \& Poli, F. 2002, ApJ,
575, 18
\bibitem[]{}
Menou, K., Haiman, Z., \& Narayan, V.K. 2001, ApJ, 558, 535
\bibitem[]{}
Monaco, P., Salucci, P., \& Danese, L. 2000, MNRAS, 311, 279
\bibitem[{Murray} 1995]{1995ApJ...451..498M} Murray, N.,
Chiang, J., Grossman, S.A., \& Voit, G.M. 1995, ApJ,  451, 498
\bibitem[{Nath} 2002]{2002MNRAS.333..145N} Nath,
B.B., \& Roychowdhury, S. 2002, MNRAS,  333, 145
\bibitem[]{}
Navarro, J.F., Frenk, C.S., \& White, S.D.M. 1997, ApJ, 490, 493
\bibitem[{Navarro} 2000]{2000ApJ...538..477N}
Navarro, J.F., \& Steinmetz, M. 2000, ApJ,  538, 477
\bibitem[]{} Peacock, J.A. 1999, Cosmological Physics (Cambridge,
Cambridge University Press)
\bibitem[\protect\citeauthoryear{O'Sullivan, Ponman, \&
Collins}{2003}]{2003MNRAS.340.1375O} O'Sullivan, E., Ponman, T.J.,
Collins, R.S. 2003, MNRAS, 340, 1375
\bibitem[{Perrotta} 2003]{2003MNRAS.338..623P}
Perrotta, F., Magliocchetti, M., Baccigalupi, C., Bartelmann, M.,
De Zotti, G., Granato, G.L., Silva, L., \& Danese, L. 2003, MNRAS,
338, 623
\bibitem[{Platania} 2002]{2002MNRAS.337..242P}
Platania, P., Burigana, C., De Zotti, G., Lazzaro, E., \&
Bersanelli, M. 2002, MNRAS,  337, 242
\bibitem[]{}
Pozzetti, L., et al. 2003, A\&A,  402, 837
\bibitem[{Press} 1974]{1974ApJ...187..425P}
Press, W.H., \& Schechter, P. 1974, ApJ,  187, 425
\bibitem[{Proga} 2000]{2000ApJ...543..686P} Proga, D.,
Stone, J.M., \& Kallman, T.R. 2000, ApJ,  543, 686
\bibitem[{Ridgway} 2002]{2002NewAR..46..175R} Ridgway, S.
Heckman, T., Calzetti, D., \& Lehnert, M. 2002, NewAR,  46, 175
\bibitem[{Romano} 2002]{2002MNRAS.334..444R} Romano, D.,
Silva, L., Matteucci, F., Danese, L. 2002, MNRAS,  334, 444
\bibitem[]{} Rawlings, S., \& Saunders, R. 1991, Nat, 349, 138
\bibitem[{Salucci} 1999]{1999MNRAS.307..637S} Salucci, P.
Szuszkiewicz, E., Monaco, P., \& Danese L. 1999, MNRAS,  307, 637
\bibitem[]{}
Sasaki, S. 1994, PASJ, 46, 427
\bibitem[{Scott} 2002]{2002MNRAS.331..817S} Scott S.E.,
Fox, M.J., Dunlop, J.S., et al. 2002, MNRAS,  331, 817
\bibitem[]{}
Sheth, R.K., et al. 2003, ApJ, submitted, astro-ph/0303092
\bibitem[]{}
Sheth, R.K., \& Tormen, G. 2002, MNRAS,  329, 61
\bibitem[{Shields} 2003]{2003ApJ...583..124S}
Shields, G.A., Gebhardt, K., Salviander, S., Wills, B.J., Xie, B.,
Brotherton, M.S., Yuan, J., \& Dietrich, M. 2003, ApJ,  583, 124
\bibitem[]{}
Silk, J., \& Rees, M.J. 1998, A\&A, 331, L1
\bibitem[]{}
Silva, L. 1999, Ph.D. Thesis, SISSA, Trieste
\bibitem[]{}
Silva, L., Granato, G.L., Bressan, A., \& Danese, L. 1998, ApJ,
509, 103
\bibitem[{Somerville} 2001]{2001MNRAS.320..504S}
Somerville, R.S., Primack, J.R., \& Faber, S.M. 2001, MNRAS,  320,
504
\bibitem[]{}
Spergel, D.N., et al. 2003, ApJ, submitted, astro-ph/0302209
\bibitem[{Stockton} 2001]{2001ApJ...554.1012S}
Stockton, A., \& Ridgway, S.E. 2001, ApJ,  554, 1012
\bibitem[]{}
Sutherland, R.S., \& Dopita, M.A. 1993, ApJS, 88, 253
\bibitem[]{}
Tavecchio, F., et al. 2000, ApJ, 543, 535
\bibitem[{Thomas} 2002]{2002Ap&SS.281..371T} Thomas, D.,
Maraston, C., \& Bender, R. 2002, Ap\&SS,  281, 371
\bibitem[{Thornton} 1998]{1998ApJ...500...95T}
Thornton, K., Gaudlitz, M., Janka, H.-T., \& Steinmetz, M. 1998,
ApJ,  500, 95
\bibitem[]{}
Trager, S.C., Faber, S.M., Worthey, G., \& Gonz\'alez, J.J. 2000a,
   AJ, 119, 1645
\bibitem[]{}
Trager, S.C., Faber, S.M., Worthey, G., \& Gonz\'alez, J.J. 2000b,
   AJ, 120, 165
\bibitem[{Tremaine} 2002]{2002ApJ...574..740T}
Tremaine, S., et al. 2002, ApJ,  574, 740
\bibitem[{Umemura} 2001]{2001ApJ...560L..29U} Umemura,
M. 2001, ApJ,  560, L29
\bibitem[{van Dokkum} 2003]{2003ApJ...587L..83V} van
Dokkum, P.G., et al. 2003, ApJ, 587, L83
\bibitem[{Valageas} 1999]{1999A&A...350..725V}
Valageas, P., \& Silk, J. 1999, A\&A,  350, 725
\bibitem[{van den Bosch} 2002]{2002MNRAS.331...98V} van
den Bosch, F.C., 2002, MNRAS,  331, 98
\bibitem[]{}
van der Marel, R.P. 1999, AJ, 117, 744
\bibitem[{Voit} 1993]{1993ApJ...413...95V} Voit, G.M.,
Weymann, R.J., \& Korista, K.T. 1993, ApJ,  413, 95
\bibitem[{Volonteri} 2002]{2002Ap&SS.281..501V}
Volonteri, M., Haardt, F., \& Madau, P. 2002, Ap\&SS,  281, 501
\bibitem[{Volonteri} 2003]{2003ApJ...582..559V}
Volonteri, M., Haardt, F., \& Madau, P. 2003, ApJ,  582, 559
\bibitem[{Wang} 1998]{1998dggn.conf..203W} Wang,
Y., \& Biermann, P.L. 1998, in Dynamics of Galaxies and Galactic
Nuclei, ed. W.J. Duschl, \& C. Einsel, 203
\bibitem[\protect\citeauthoryear{Wada \& Venkatesan}{2003}]
{2003ApJ...591...38W} Wada, K., \& Venkatesan,
A. 2003, ApJ, 591, 38
\bibitem[{Wechsler} 2002]{2002ApJ...568...52W}
Wechsler, R.H., Bullock, J.S., Primack, J.R., Kravtsov, A.V., \&
Dekel, A. 2002, ApJ, 568, 52
\bibitem[{Woosley} 1986]{1986ARA&A..24..205W}
Woosley, S.E., \& Weaver, T.A. 1986, ARA\&A,  24, 205
\bibitem[\protect\citeauthoryear{Worthey et
al.}{1994}]{1994ApJS...94..687W} Worthey, G., Faber, S.M.,
Gonzalez, J.J., \& Burstein, D. 1994, ApJS, 94, 687
\bibitem[]{}
Wu, K.K.S., Fabian, A.C., \& Nulsen, P.E.J. 2000, MNRAS, 318, 889
\bibitem[{Zhao} 2003]{2003MNRAS.339...12Z} Zhao, D.H.,
Mo, H.J., Jing, Y.P., \& B{\" o}rner, G. 2003, MNRAS,  339, 12
\bibitem[{Zepf} 1996]{1996ApJ...466..114Z} Zepf,
S.E., \&  Silk, J. 1996, ApJ,  466, 114
\end{thebibliography}
\end{document}